\documentclass[10pt,twocolumn,letterpaper]{article}

\usepackage{iccv}
\usepackage{times}
\usepackage{epsfig}
\usepackage{graphicx}
\usepackage{amsmath}
\usepackage{amssymb}
\usepackage{multirow}
\usepackage{subfigure}
\usepackage[dvipsnames]{xcolor}
\usepackage[linesnumbered,ruled,vlined]{algorithm2e}

\graphicspath{ {images/} {images-suppl/} }

\newcommand\norm[1]{\left\lVert#1\right\rVert}

\usepackage[pagebackref=true,breaklinks=true,letterpaper=true,colorlinks,bookmarks=false]{hyperref}

\iccvfinalcopy 

\def\iccvPaperID{1976} 

\ificcvfinal\fi

\begin{document}

\title{Collaborative Unsupervised Visual Representation Learning \protect\\ from Decentralized Data}

\author{Weiming Zhuang$^{1, 3}$ \quad Xin Gan$^{2}$ \quad Yonggang Wen$^{2}$ \quad Shuai Zhang$^{3}$ \quad Shuai Yi$^{3}$\\
$^{1}$S-Lab, Nanyang Technological University, $^{2}$Nanyang Technological University, 
$^{3}$SenseTime Research \\
{\tt\small \{weiming001,ganx0005\}@e.ntu.edu.sg,ygwen@ntu.edu.sg,\{zhangshuai,yishuai\}@sensetime.com}
}

\maketitle
\ificcvfinal\thispagestyle{empty}\fi

\begin{abstract}

Unsupervised representation learning has achieved outstanding performances using centralized data available on the Internet. However, the increasing awareness of privacy protection limits sharing of decentralized unlabeled image data that grows explosively in multiple parties (e.g., mobile phones and cameras). As such, a natural problem is how to leverage these data to learn visual representations for downstream tasks while preserving data privacy. To address this problem, we propose a novel federated unsupervised learning framework, \textit{FedU}. In this framework, each party trains models from unlabeled data independently using contrastive learning with an online network and a target network. Then, a central server aggregates trained models and updates clients' models with the aggregated model. It preserves data privacy as each party only has access to its raw data. Decentralized data among multiple parties are normally non-independent and identically distributed (non-IID), leading to performance degradation. To tackle this challenge, we propose two simple but effective methods: 1) We design the communication protocol to upload only the encoders of online networks for server aggregation and update them with the aggregated encoder; 2) We introduce a new module to dynamically decide how to update predictors based on the divergence caused by non-IID. The predictor is the other component of the online network. Extensive experiments and ablations demonstrate the effectiveness and significance of FedU. It outperforms training with only one party by over 5\% and other methods by over 14\% in linear and semi-supervised evaluation on non-IID data. 

\end{abstract}


\section{Introduction}

Learning good visual representations without supervision is attracting considerable attention in recent years. These visual representations can facilitate efficient training of downstream tasks \cite{simonyan2014very-deep-cnn, oquab2014transferring-downstream} like image segmentation \cite{long2015segmentation}. Researchers have proposed many unsupervised representation learning methods by designing pretext tasks \cite{doersch2015unsup-context-prediction, zhang2016colorful-image, noroozi2016jigsaw, gidaris2018unsup-image-rotation}. Among them, contrastive learning \cite{hadsell2006constrastive-loss, oord2018contrastive-predictive-coding} based on instance-level discrimination has achieved the state-of-the-art performance \cite{wu2018unsupervised-instance, chen2020simclr, he2020moco, grill2020byol, chen2020simsiam}. These unsupervised representation learning methods rely on the assumption that data can be collected and stored in a centralized database, such as images from the Internet.

However, in real-world scenarios, decentralized image data are growing explosively and the data collected in multiple parties may not be centralized due to data privacy regulations \cite{gdpr}. For example, photos taken on phones or images collected from street cameras could contain sensitive information. Centralizing them for training could reveal the identity and locality of individuals \cite{zhuang2020fedreid}. Besides, compared with training representations using only publicly available data from the Internet, representations learned from the actual data may be more representative for downstream tasks applied on similar scenarios. Utilizing the decentralized unlabeled image data to learn representations with privacy guaranteed is an important but overlooked problem.

\begin{figure}[t]
\begin{center}
  \subfigure[Single Client Training]{\label{fig:tsne-single}\includegraphics[width=38mm]{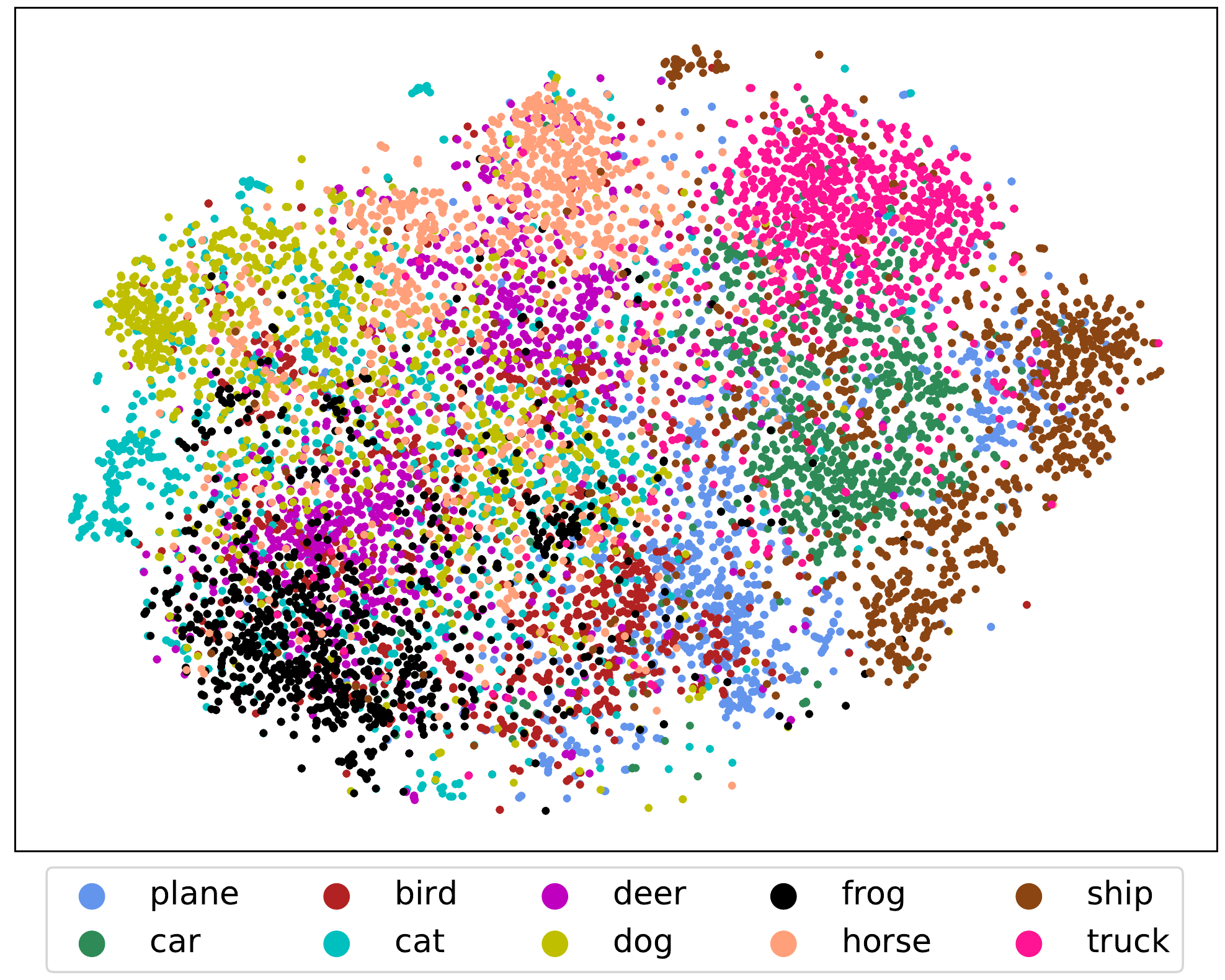}}
  \subfigure[Our Proposed FedU]{\label{fig:tsne-fedu-}\includegraphics[width=38mm]{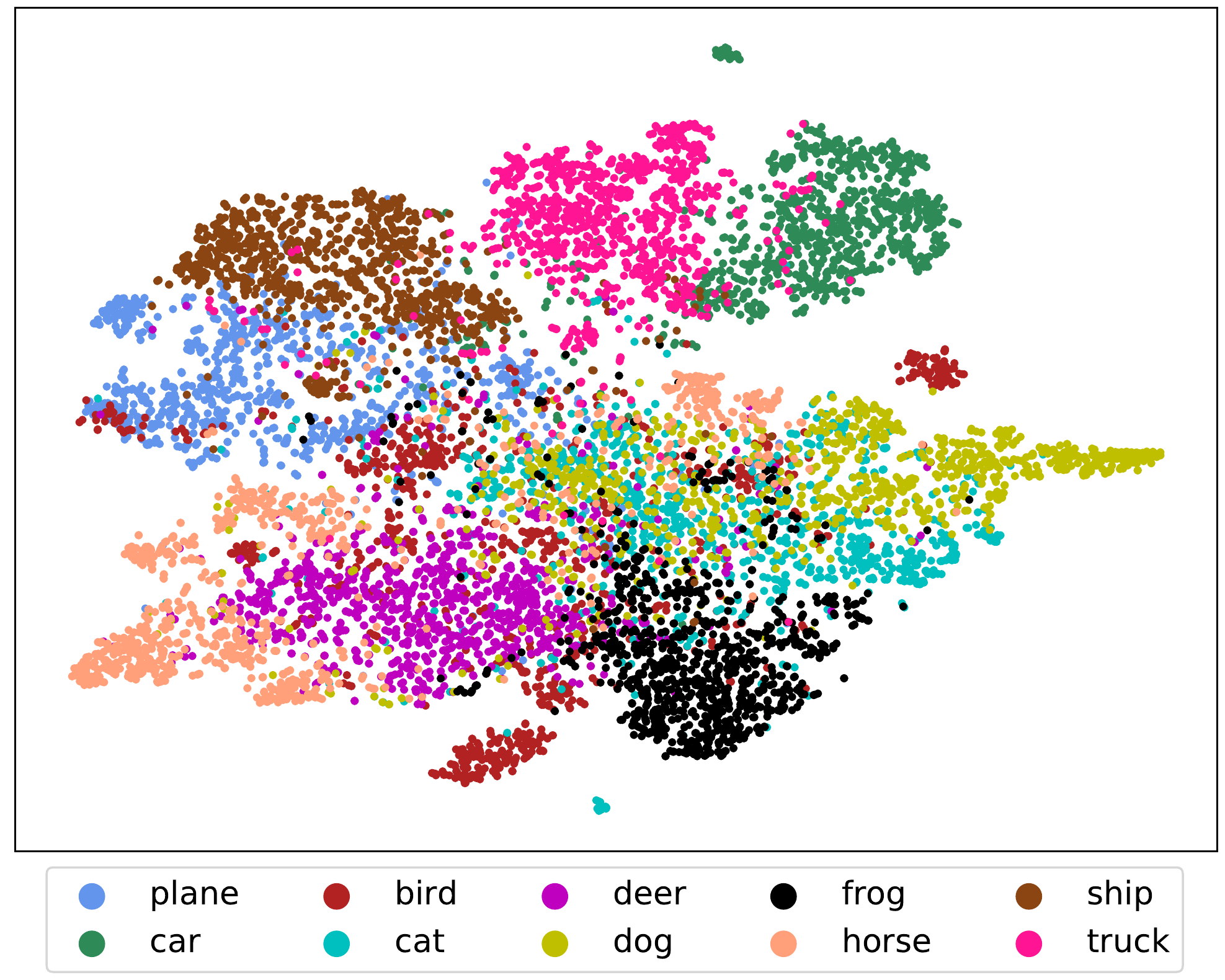}}
\end{center}
  \caption{Limitations of unsupervised learning (BYOL \cite{grill2020byol}) on non-IID data. (a) and (b) are t-SNE visualization of representations. We simulate five decentralized clients, where each client has unique two out of ten classes of CIFAR-10 \cite{cifar10-2009} training data. Training in a single client cannot learn the representation properly.}
  \label{fig:tsne-motivation}
\end{figure}

Existing methods cannot leverage decentralized unlabeled data to learn a generic representation while preserving data privacy. Federated learning is an emerging distributed training technique \cite{fedavg} that have empowered multiple parties to collaboratively learn computer vision models, such as segmentation \cite{li2019brain-tumor1, sheller2018brain-tumor2}, object detection \cite{luo2019real-obj-detect}, and person re-identification \cite{zhuang2020fedreid}. But traditional federated learning relies on data labels. Decentralized data is normally non-independent and identically distributed (non-IID), which is one of the key challenges for learning from multiple parties \cite{zhao2018non-iid, Li2020FedChallenges}. For example, in real-world street datasets \cite{luo2019real-obj-detect}, cameras capture diverse two or three out of seven object categories. The state-of-the-art unsupervised learning approaches are effective, but they may not work well with non-IID data, as shown in Figure \ref{fig:tsne-single}. Although Federated Contrastive Averaging (FedCA) \cite{zhang2020fedca} addresses the unlabeled and non-IID problems, it imposes potential privacy leakage risks by directly sharing features of clients' data.

In this paper, we propose a new federated unsupervised representation learning framework, \textit{FedU}, to learn generic visual representations collaboratively from decentralized unlabeled data in multiple parties while preserving data privacy. Built on contrastive learning for unlabeled data in each party \cite{grill2020byol}, we propose to centralize the learned representations instead of raw data to a central server to protect data privacy. Each party uses Siamese networks in training: an online network with an online encoder and a predictor; a target network with a target encoder. FedU is not a trivial combination of contrastive learning and federated learning, because we implement two simple but effective methods to tackle the non-IID data problem. As studied in \cite{zhao2018non-iid}, non-IID data causes weight divergence, resulting in performance degradation. Through analyzing the characteristics of the Siamese networks and the impact of non-IID data, we first design a communication protocol to upload only the online encoders of parties for server aggregation and update them with the aggregated global encoder for the next training round. Moreover, we introduce a new module, divergence-aware predictor update (DAPU), to dynamically decide the choice of predictor update based on the degree of divergence. It updates the predictors in multiple parties with the aggregated predictor only when the divergence is smaller than a threshold.

We evaluate FedU on CIFAR datasets \cite{cifar10-2009} using various backbones and settings. Extensive experiments demonstrate that FedU achieves promising results under three evaluation protocols: linear evaluation, semi-supervised learning, and transfer learning. Compared with existing methods, FedU achieves superior performance on all evaluation protocols. Specifically, the representation learned from FedU is much better than the one learned from only one party (Figure \ref{fig:tsne-motivation}). It outperforms training with one party by over 5\% and other methods by more than 14\% under linear and semi-supervised evaluations on non-IID data. We also present ablations to illustrate the intuition and performance of FedU.


In summary, the contributions of the paper are:

\begin{itemize}

    \item We introduce a new framework, \textit{FedU}, to address an important but overlooked problem: leveraging unlabeled data from multiple parties to learn visual representations while preserving data privacy.
    
    \item We design the communication protocol to only aggregate and update the online encoders by analyzing the impact of non-IID data on Siamese networks.
    
    \item We propose a new module, divergence-aware predictor update (DAPU), to dynamically determine how to update predictors based on the degree of divergence caused by non-IID data.
    
    
    \item Extensive experiments and ablations demonstrate the effectiveness and significance of FedU. 
\end{itemize}

\begin{figure*}[t]
\begin{center}
\includegraphics[width=0.95\linewidth]{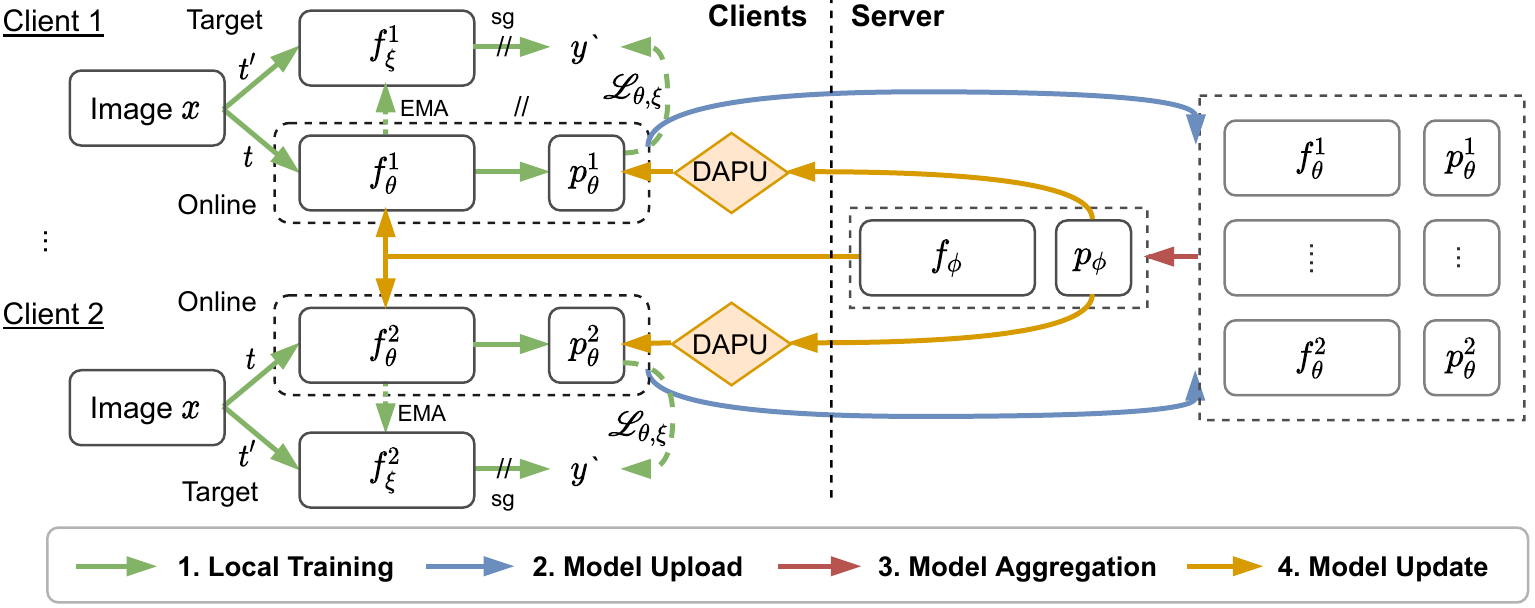}
   \caption{Overview of our proposed federated unsupervised representation framework (FedU) that trains representations collectively with multiple clients coordinated by a central server. Each training round of FedU consists of four stages: (1) \textbf{Local training}: each client $k$ trains an online encoder $f_\theta^k$ and a predictor $p_\theta^k$ with contrastive loss and updates a target encoder $f_\xi^k$ with exponential moving average (EMA). (sg means stop gradient.) (2) \textbf{Model Upload}: Each client $k$ uploads $f_\theta^k$ and $p_\theta^k$ to the server. (3) \textbf{Model Aggregation}: The server aggregates all $f_\theta^k$ and $p_\theta^k$ to obtain a new global encoder $f_\phi$ and predictor $p_\phi$. (4) \textbf{Model Update}: The server distributes and updates clients' local encoders $f_\theta$ with $f_\phi$. The clients update their predictors $p_\theta$ based on our proposed divergence-aware predictor update (DAPU).}
\label{fig:architecture}
\end{center}
\end{figure*}




\section{Related Work}

\subsection{Unsupervised Representation Learning}

The majority of unsupervised representation learning methods fall into two categories: generative and discriminative. Generative methods learn representations by generating pixels mapping to the input space via methods like auto-encoding \cite{vincent2008autoencoder, kingma2013auto-encoding} or adversarial learning \cite{NIPS2014GAN, radford2015adversarial}. Discriminative methods learn representations by performing proxy tasks, such as image in-painting \cite{pathak2016inpainting} and solving jigsaw puzzles \cite{noroozi2016jigsaw}. Among discriminative approaches, contrastive learning is the state-of-the-art method \cite{wu2018unsupervised-instance, chen2020simclr, he2020moco, grill2020byol, chen2020simsiam}. Contrastive learning \cite{hadsell2006constrastive-loss, oord2018contrastive-predictive-coding} aims to minimize the similarity of positive samples (two different augmentations of the same image), while maximizing the similarity of negative samples (two different images). The negative pairs are either generated from a memory bank like MoCo \cite{he2020moco} or from a large batch size like SimCLR \cite{chen2020simclr}. Methods like BYOL \cite{grill2020byol} and SimSiam \cite{chen2020simsiam} even bypass negative pairs and contrast only positive pairs. However, these methods do not perform well with non-IID data. In our proposed FedU, we introduce two methods to tackle the non-IID data challenge.



\subsection{Federated Learning}

Federated Learning (FL) is an emerging distributed training method that coordinates decentralized clients to train machine learning models \cite{fedavg}. FedAvg \cite{fedavg} is the standard algorithm for FL. Non-IID data is one of the key challenges of FL, which causes weight divergence (Figure \ref{fig:divergence}) and performance drop as discussed in \cite{zhao2018non-iid}. Many approaches have been proposed to address this challenge, such as sharing a public dataset \cite{zhao2018non-iid}, knowledge distillation \cite{zhuang2020fedreid}, or regularizing training in clients \cite{fedprox}. However, researchers investigate these methods under supervised learning, not directly applicable to scenarios where data is unlabeled.



Although several existing works study federated learning with unlabeled data \cite{zhuang2021fedfr, zhuang2021fedureid}, they mainly focus on specific applications. Other methods either bypass the non-IID problem \cite{van2020fedae, jin2020fed-unlabeled-survey} or impose potential privacy risks \cite{zhang2020fedca}. Specifically, FedCA \cite{zhang2020fedca} solves the non-IID problem by gathering features and data distribution from clients, compromising data privacy from potential attacks. Our proposed FedU is simpler, achieves better performance, and effectively preserves data privacy.

\section{Methodology}
\label{sec:methodology}

In this section, we first define the problem and then introduce our proposed federated unsupervised representation learning framework (FedU) to address the problems.

\subsection{Problem Definition}

Before presenting the details of FedU, we define the problem and the assumptions first. Several parties aim to learn a generic representation $f_\phi$ for various downstream tasks without sharing data among these parties. We denote each party as a client $k$ containing unlabeled data $\mathcal{D}_k = \{\mathcal{X}_k\}$. The global objective function is $\min_{\phi} h(\phi) :=  \sum_{k=1}^N p_k F_k(\phi)$, where $N$ is the number of clients, $p_k = \frac{n_k}{n}$, and $n = \sum_k n_k$ is the total data size. For client $k$, $F_k(\phi) := \mathbb{E}_{x_k \sim \mathcal{D}_k}[h_k(\phi;x_k)]$ is the expected loss over data distribution $\mathcal{D}_k$, where $x_k$ is the data and $h_k(\phi;x_k)$ represents the loss function to train models. 

\begin{figure}[t]
\begin{center}
\includegraphics[width=0.7\linewidth]{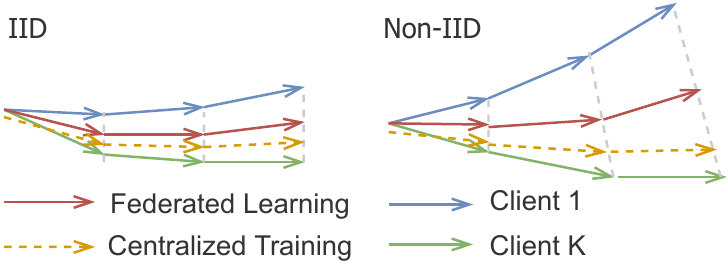}
\end{center}
  \caption{Illustration of weight divergence caused by non-IID data studied in \cite{zhao2018non-iid}. Federated learning on non-IID data leads to significant divergence from the centralized training.}
\label{fig:divergence}
\end{figure}

A key challenge is that data among decentralized clients are likely to be non-IID. For example, each client could contain only two out of seven object categories in real-world street datasets \cite{luo2019real-obj-detect}. As discussed in \cite{zhao2018non-iid}, non-IID data causes weight divergence (illustrated in Figure \ref{fig:divergence}). Training with non-IID data in one client using existing unsupervised learning methods could result in a poor representation, as shown in Figure \ref{fig:tsne-single}. Besides, we cannot centralize data from clients to construct a big dataset due to privacy constraints. In this paper, we aim to leverage the growing amount of unlabeled image data from multiple parties to learn a generic representation $f_\phi$ without privacy leakage.

\subsection{FedU Overview}

Figure \ref{fig:architecture} presents the overview of our proposed framework, FedU, which tackles the problem defined above. FedU instructs a server to coordinate multiple clients with unlabeled data to train a generic representation $f_\phi$. It follows the proposed communication protocol to upload the online encoder $f_\theta$ for server aggregation and update them with the global encoder $f_\phi$. Besides, FedU introduces a new divergence-aware predictor update (DAPU) module to address the non-IID data challenge.

Before presenting technical details, we introduce each round's training pipeline of FedU with following stages: 
(1) \textit{Local training}: each client $k$ conducts unsupervised representation learning with contrastive loss (Equation \ref{eq:loss}), obtaining an online network with an online encoder $f_\theta^k$ and a predictor $p_\theta^k$, as well as a target network with a target encoder $f_\xi^k$. 
(2) \textit{Model Upload}: Each client $k$ uploads the online encoder $f_\theta^k$ and the predictor $p_\theta^k$ to the server. 
(3) \textit{Model Aggregation}: The server aggregates clients' online encoders and predictors to obtain a new global encoder with $f_\phi = \sum_{k=1}^N \frac{n_k}{n} f_\theta^k$ and a new global predictor with $p_\phi = \sum_{k=1}^N \frac{n_k}{n} p_\theta^k$, where $n_k$ is the data volume of client $k$ and $n$ is the total data volume of $N$ clients. 
(4) \textit{Model Update}: The server sends the global encoder $f_\phi$ and predictor $p_\phi$ to all clients. Each client updates its online encoder $f_\theta$ with the global encoder $f_\phi$ and uses DAPU to dynamically decide whether updating local predictor $p_\theta$ with the global predictor $p_\phi$ (Equation \ref{eq:dapu}). We summarize FedU in Algorithm \ref{algo:fedu}. Next, we present the details of local training, the communication protocol, and DAPU.




\subsection{Local Training}
\label{sec:local-training}

In local training, each client conducts contrastive learning with asymmetric Siamese networks: an online network and a target network, adopted from BYOL \cite{grill2020byol}. Traditional federated learning only needs one network in each client to perform supervised training. However, as data is unlabeled, FedU requires two networks to generate positive pairs from two augmentations of an image for contrastive learning. The online network consists of an online encoder $f_\theta$ and a predictor $p_\theta$. The target network only contains a target encoder $f_\xi$. $f_\theta$ and $f_\xi$ share the same architecture but differ in parameters. 

Local training starts from taking two augmentations $t$ and $t^{'}$ of the input image $x$. They are the inputs of the online and target networks, respectively. The role of the online network is to learn from the unlabeled data. We update its parameters $\theta$ with contrastive loss:
\begin{equation}
    \mathcal{L}_{\theta, \xi} \triangleq \norm{y - y^{'}}^2_2 \triangleq 2 - 2 \cdot \frac{\langle y, y^{'} \rangle}{\norm{y}_2 \cdot \norm{y^{'}}_2},
\label{eq:loss}
\end{equation}
where $y \triangleq p_\theta(f_\theta(t))$ is the output of the online network and $y^{'} \triangleq f_\xi(t^{'})$ is the output of the target network. The role of the target network is to generate a positive regression target for the online network to contrast. Instead of updating with gradient descent, its parameters $\xi$ are updated with the exponential moving average (EMA) of the online encoder's parameters $\theta$ in every batch:
\begin{equation}
    \xi = m\xi + (1-m)\theta, 
\label{eq:target}
\end{equation}
where $m \in [0, 1]$ is the decay rate. 

Each client trains $E$ local epochs and then uploads model updates to the server. The communication protocol between the server and clients requires careful considerations to mitigate the adverse impact of non-IID data.

\SetKwInput{KwInput}{Input}                
\SetKwInput{KwOutput}{Output}              
\SetKwFunction{FnClient}{}
\SetKwFunction{FnServer}{}

\begin{algorithm}[t]
    \caption{Federated Unsupervised Representation Learning Framework (FedU)}
    \label{algo:fedu}
    \SetAlgoLined
    \KwInput{learning rate $\eta$, threshold $\mu$, total training rounds $R$, $E, B, K, N, n_k, n$}
    \KwOutput{$f_\phi$}
    
    \SetKwProg{Fn}{Server}{:}{}
    \Fn{\FnServer}{
        Initialize the global encoder $f_\phi^0$ and predictor $p_\phi^0$\;
        \For{each round r = 0 to R-1}{
            $S_t \leftarrow$ (random selection of K clients)\;
            \For{client $k \in S_t$ concurrently}{
                $f_\theta^k, p_\theta^k \leftarrow $ \textbf{Client}($f_\phi^r$, $p_\phi^r$, $r$)\;
            }
            \textcolor{BlueViolet}{// Model aggregation}\;
            $f_\phi^{r+1} \leftarrow \sum_{k \in S_t} \frac{n_k}{n} f_\theta^k$\;
            $p_\phi^{r+1} \leftarrow \sum_{k \in S_t} \frac{n_k}{n} p_\theta^k$\;
        }
        \KwRet $f_\phi^{R}$\;
    }
    
    \SetKwProg{Fn}{Client}{:}{\KwRet}
    \Fn{\FnClient{$f_\phi$, $p_\phi$, $r$}}{
        \textcolor{BlueViolet}{// Model update}\;
        $f_\theta \leftarrow f_\phi$\;
        \textcolor{BlueViolet}{// Update using DAPU with Equation \ref{eq:dapu}}\;
        \If{$\norm{\theta^{r} - \phi^{r-1}}_2^2 < \mu$}{
            $p_\theta \leftarrow p_\phi$\;
        }
        $\mathcal{Z} \leftarrow $ (construct data batches with batch size $B$)\;
        \textcolor{BlueViolet}{// Local training with Equation \ref{eq:loss} and \ref{eq:target}}\;
        \For{local epoch e = 0 to E-1}{
            \For{$z \in \mathcal{Z}$}{
                $\theta \leftarrow \theta - \eta \triangledown \mathcal{L_{\theta, \xi}}(\theta; z)$\;
                $\xi \leftarrow m\xi + (1-m)\theta$\;
            }
        }
        \KwRet $f_\theta$, $p_\theta$ \textcolor{BlueViolet}{// Model upload}\; 
    }
\end{algorithm}

\subsection{Communication Protocol}
\label{sec:encoder}

FedU requires bidirectional communication between the server and the clients. In the model upload stage, clients send models to the server for aggregation. In the model update stage, the server distributes the aggregated model to clients and updates clients' local models. As FedU performs local training with two models of the same architecture, it leads to important design decisions on the communication protocol: (1) Which encoder (online or target) to \textit{upload} for aggregation? (2) Which encoder (online, target, or both) to \textit{update} with the aggregated encoder from the server?


We analyze the characteristics of both encoders and hypothesize that it is desirable to aggregate and update the \textit{online encoder}. The target encoder is the exponential averaging of the online encoder, denoting the historical representation of a client. The online encoder is constantly updated from back-propagation in every training step, representing the latest representation of a client. Although both encoders capture local data characteristics, we argue that the latest representation learned in the online encoder is more representative of local data distribution. Thus, uploading the online encoder for server aggregation can better capture the characteristics of non-IID data. 

In the model update stage, the online encoder also plays an important role. The server aggregated global encoder $f_\phi$ is the average of encoders from clients, so it has better generalizability. As the target encoder produces regression targets for the online encoder to contrast, we should not only update the target encoder with a more general global model $f_\phi$ because it would decrease its capability of providing local-representative targets. Updating both the online and target encoders could work, but it means that the local training of each client needs to adapt the general global model $f_\phi$ to non-IID local data again. Hence, we only update the online encoder $f_\theta$ with the global encoder $f_\phi$ while keeping the parameters of the target encoder for stable regression targets. Our ablation study of the online and target encoders (Table \ref{tab:abs-encoders}) validates our hypothesis --- using the online encoder for aggregation and update.

\begin{table*}[t]
\begin{center}
\begin{tabular}{llcccccc}
\hline
\multicolumn{1}{l}{\multirow{2}{*}{Method}} &
\multicolumn{1}{l}{\multirow{2}{*}{Architecture}} &
\multicolumn{1}{c}{\multirow{2}{*}{Param.}} &
\multicolumn{2}{c}{CIFAR-10} &
\multicolumn{1}{c}{} &
\multicolumn{2}{c}{CIFAR-100}
\\ 
\cline{4-5} \cline{7-8}
\multicolumn{1}{c}{} & & & IID & Non-IID & & IID & Non-IID
\\
\hline 
\hline 

Single client training & ResNet-18 & 11M & 81.24 & 71.98 & & 51.33 & 49.69 \\

Single client training & ResNet-50 & 23M & 83.16 & 77.84 & & 57.21 & 55.16 \\
FedSimCLR \cite{chen2020simclr} \cite{zhang2020fedca} & ResNet-50 & 23M & 68.10 & 64.06 & & 39.75 & 38.70 \\ 
FedCA \cite{zhang2020fedca}  & ResNet-50 & 23M & 71.25 & 68.01 & & 43.30 & 42.34 \\ 
FedSimSiam \cite{chen2020simsiam} & ResNet-50 & 23M & 79.64 & 76.70 & & 46.28 & 48.80 \\
FedU (ours) & ResNet-18 & 11M & 85.21 & 78.71 & & 56.52 & 57.08 \\ 
FedU (ours) & ResNet-50 & 23M & \textbf{86.48} & \textbf{83.25} & & \textbf{59.51} & \textbf{61.94} \\ 

\hline
\multicolumn{8}{l}{\textit{Upper-bound methods: centralized unsupervised learning and supervised federated learning}} \\
\hline
FedAvg \cite{fedavg} (Supervised) & ResNet-50 & 23M & 91.51 & 67.74 & & 65.77 & 64.38 \\
BYOL \cite{grill2020byol} (Centralized) & ResNet-50 & 23M & 91.85 & - & & 66.51 & - \\

\hline
\end{tabular}
\end{center}
\caption{Top-1 accuracy (\%) comparison under the linear evaluation protocol on IID and non-IID settings of CIFAR datasets. Our proposed FedU outperforms other methods. It even outperforms supervised federated learning (FedAvg) on the non-IID setting of CIFAR-10 dataset.}
\label{tab:linear-eval}
\end{table*}

\begingroup
\setlength{\tabcolsep}{0.5em}
\begin{table*}[t]
\begin{center}
\begin{tabular}{llcccccccccccc}
\hline
\multicolumn{1}{l}{\multirow{3}{*}{Method}} &
\multicolumn{1}{l}{\multirow{3}{*}{Architecture}} &
\multicolumn{1}{c}{\multirow{3}{*}{Param.}} &
\multicolumn{5}{c}{CIFAR-10} &
\multicolumn{1}{c}{} &
\multicolumn{5}{c}{CIFAR-100}
\\ 
\cline{4-8} \cline{10-14}
\multicolumn{1}{c}{} & & & \multicolumn{2}{c}{IID} & &  \multicolumn{2}{c}{Non-IID} & & \multicolumn{2}{c}{IID} & &  \multicolumn{2}{c}{Non-IID}
\\
\cline{4-5} \cline{7-8} \cline{10-11} \cline{13-14}
\multicolumn{1}{c}{} & & & 1\% & 10\% & & 1\% & 10\% & & 1\% & 10\% & & 1\% & 10\%
\\
\hline 
\hline 

Single client training & ResNet-18 & 11M & 74.76 & 78.08 & & 60.25 & 70.60 & & 26.32 & 43.05 & & 21.95 & 37.70 \\

Single client training & ResNet-50 & 23M & 74.80 & 80.33 & & 63.65 & 74.30 & & 25.91 & 45.29 & & 23.18 & 41.43 
\\
FedAvg \cite{fedavg} (Super.) & ResNet-50 & 23M & 26.68 & 40.44 & & 17.72 & 21.69 & & 8.09 & 5.37 & & 14.47 & 13.98 
\\ 
FedSimCLR \cite{chen2020simclr} \cite{zhang2020fedca} & ResNet-50 & 23M & 50.00 & 60.67 & & 26.03 & 33.83 & & 23.01 & 31.56 & & 14.02 & 20.01
\\
FedCA \cite{zhang2020fedca} & ResNet-50 & 23M & 50.67 & 61.02 & & 28.50 & 36.28 & & 23.32 & 32.09 & & 16.48 & 22.46 
\\ 
FedU (ours) & ResNet-18 & 11M & 79.40 & 82.61 & & 68.28 & 78.52 & & \textbf{31.31} & \textbf{47.64} & & 30.36 & \textbf{48.80} 
\\
FedU (ours) & ResNet-50 & 23M & \textbf{79.44} & \textbf{83.08} & & \textbf{71.19} & \textbf{80.08} & & 29.35 & 47.14 & & \textbf{30.80} & 48.76 
\\
\hline 
BYOL \cite{grill2020byol} (Centralized) & ResNet-50 & 23M & 89.07 & 89.66 & & - & - & & 41.49 & 60.23 & & - & - 
\\ 
\hline
\end{tabular}
\end{center}
\caption{Top-1 accuracy (\%) comparison under semi-supervised protocol using 1\% and 10\% of CIFAR datasets for fine-tuning. FedU outperforms other methods except the upper-bound --- centralized unsupervised learning (BYOL).}
\label{tab:semi-sup}
\end{table*}    
\endgroup

\subsection{Divergence-aware Predictor Update}
\label{sec:predictor}

Apart from considering the encoders for the communication protocol, another vital design choice is whether clients should update the predictors with the global predictor in every training round. Inspired by \cite{zhao2018non-iid} that non-IID data causes weight divergence (Figure \ref{fig:divergence}), we propose a new module, divergence-aware predictor update (DAPU), to dynamically decide whether updating the local predictor $p_\theta$ with the aggregated predictor $p_\phi$. We make the decision based on the degree of divergence and formulate it as:

\begin{equation}
    p_{\theta} =
    \begin{cases}
      p_{\phi} & \norm{\theta^{r} - \phi^{r-1}}_2^2 < \mu \\
      p_{\theta} & \text{otherwise}
    \end{cases}
\label{eq:dapu}
\end{equation}

where $\mu$ is a controllable threshold. $\theta^r$ and $\phi^{r-1}$ represents the parameters of the online encoder in round $r$ and the global encoder in round $r-1$ respectively. As local encoder parameters $\theta^{r-1}$ is updated with the global encoder parameters $\phi^{r-1}$, $\norm{\theta^{r} - \phi^{r-1}}_2^2$ measures the divergence of model parameters occurred in local training. 

The intuition of DAPU is that clients update the predictors with the local predictor $p_\theta$ when divergence is large and update them with the global predictor $p_\phi$ when divergence is small. The predictor is the last layer of the online network. Based on the study of characteristics of layers of convolutional neural network \cite{zeiler2014visualizing-cnn}, the last layer captures information that is most related to specific classes and objects in the dataset. Since local data in clients is non-IID, the predictors $p_\theta$ of clients could have large variances. Simply updating predictors in clients with the global predictor $p_\phi$ could have side effects on learning. While on the other hand, always updating with local predictors limits clients' generalizability. Hence, we propose to dynamically update it with global predictor $p_\phi$ only when the divergence is small.

\section{Experimental Evaluation}
\label{sec:experiments}

In this section, we evaluate the performance of the representation $f_\phi$ learned by FedU on CIFAR-10 and CIFAR-100 \cite{cifar10-2009}. We first explain the experiment setup. Then we assess the representation in linear evaluation, semi-supervised learning, and transfer capabilities to other datasets. 


\subsection{Experiment Setup}

\textbf{Datasets and Federated Simulations} For linear and semi-supervised evaluation, we use CIFAR-10 and CIFAR-100 \cite{cifar10-2009} datasets. Both contain 50,000 training images and 10,000 testing images. CIFAR-10 and CIFAR-100 contain 10 classes and 100 classes with an equal number of images per class, respectively. For transfer learning evaluation, we train on the Mini-ImageNet dataset \cite{vinyals2016mini-imagenet}. Mini-ImageNet contains 60,000 images in 100 classes extracted from ImageNet \cite{imagenet_cvpr09}. 
To simulate $K$ clients, we divide training sets into $K$ partitions. For IID simulation, each client contains an equal number of images of all classes. For non-IID simulation, each client contains $\frac{10}{K}$ CIFAR-10 classes and $\frac{100}{K}$ CIFAR-100/Mini-ImageNet classes.





\textbf{Implementation Details} We implement FedU in Python using EasyFL \cite{zhuang2021easyfl} based on PyTorch \cite{paszke2017pytorch} framework. We simulate $K$ clients of training using $K$ NVIDIA\textsuperscript{\textregistered} V100 GPUs, one GPU for one client. The server and clients communicate through PyTorch communication backend. We use ResNet-18 and ResNet-50 \cite{he2016resnet} as the network architecture for the encoders and use a multi-layer perceptron (MLP) as the predictor. For fair comparison with other methods, we run experiments with $K = 5$ clients for $R = 100$ training rounds, where each client performs $E = 5$ local epochs in each round. We use the threshold $\mu = 0.4$ and $\mu = 0.6$ for CIFAR-10 and CIFAR-100 experiments, respectively. For ablation studies, each client conducts $E = 1$ local epochs in each round for $R = 800$ rounds. We use decay rate $m = 0.99$, batch size $B = 128$, and SGD as optimizer with learning rate $\eta = 0.032$.


\subsection{Linear Evaluation}

We evaluate the representation learned from FedU using linear evaluation on CIFAR datasets, following the linear evaluation protocol described in \cite{kolesnikov2019revisiting, grill2020byol}: We first train a model without supervision using FedU and other baseline methods for 100 epochs; Next, we froze the model parameters of the backbone and train a new classifier on top of it for another 100 epochs. The following are the compared methods: (1) \textit{Single Client Training}: each client learns a representation with their local data for 500 epochs using BYOL \cite{grill2020byol}; (2) \textit{FedSimCLR}: simply combine federated learning and SimCLR \cite{chen2020simclr} from paper \cite{zhang2020fedca}; (3) \textit{FedSimSiam}: combine federated learning with SimSiam \cite{chen2020simsiam} (use SimSiam for local training instead of our method); (4) \textit{FedCA}: the method proposed in \cite{zhang2020fedca}. All the experiments are conducted under the same settings. Besides, we also compare FedU with two potential upper bound methods: centralized unsupervised learning using BYOL \cite{grill2020byol} and supervised federated learning with FedAvg \cite{fedavg}. 

Table \ref{tab:linear-eval} reports the performance of these methods with different backbones, datasets, and under both IID and non-IID settings. It shows that FedU achieves better performance than other methods on linear evaluation. Specifically, it outperforms the existing method FedCA \cite{zhang2020fedca} by at least 14\% and outperforms the single client training by more than 5\% on non-IID settings. Compared with the theoretical upper-bound methods BYOL and FedAvg, FedU outperforms them on non-IID CIFAR-10 data. Besides, the results show that the performance of FedU increases using deeper backbones (ResNet-50 vs. ResNet-18 \cite{he2016resnet}).

\begin{figure*}[t]
\begin{center}
  \subfigure[Agg. Online, Update Target]{\label{fig:online-target}\includegraphics[width=37mm]{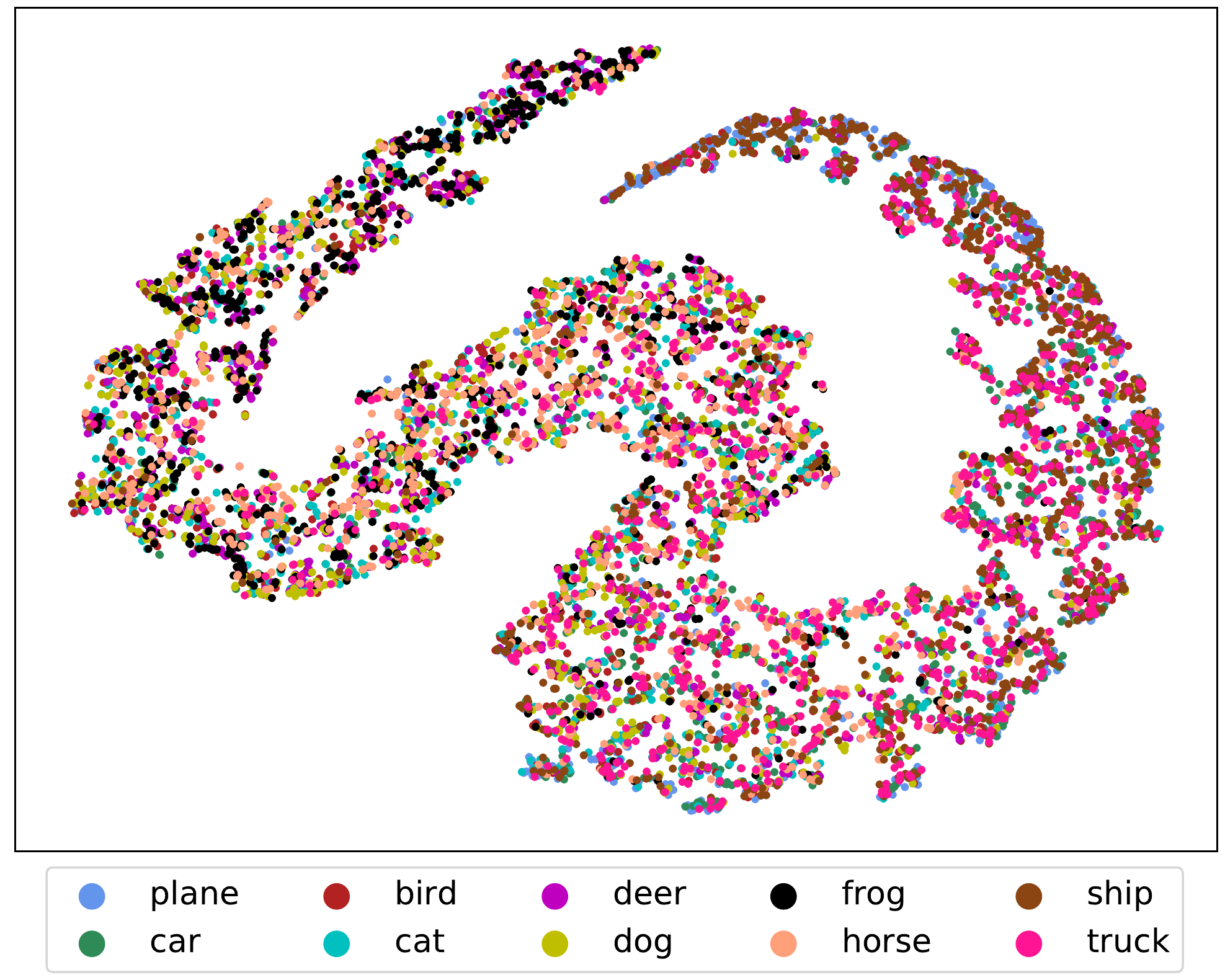}}
  \subfigure[Agg. Target, Update Online]{\label{fig:target-online}\includegraphics[width=37mm]{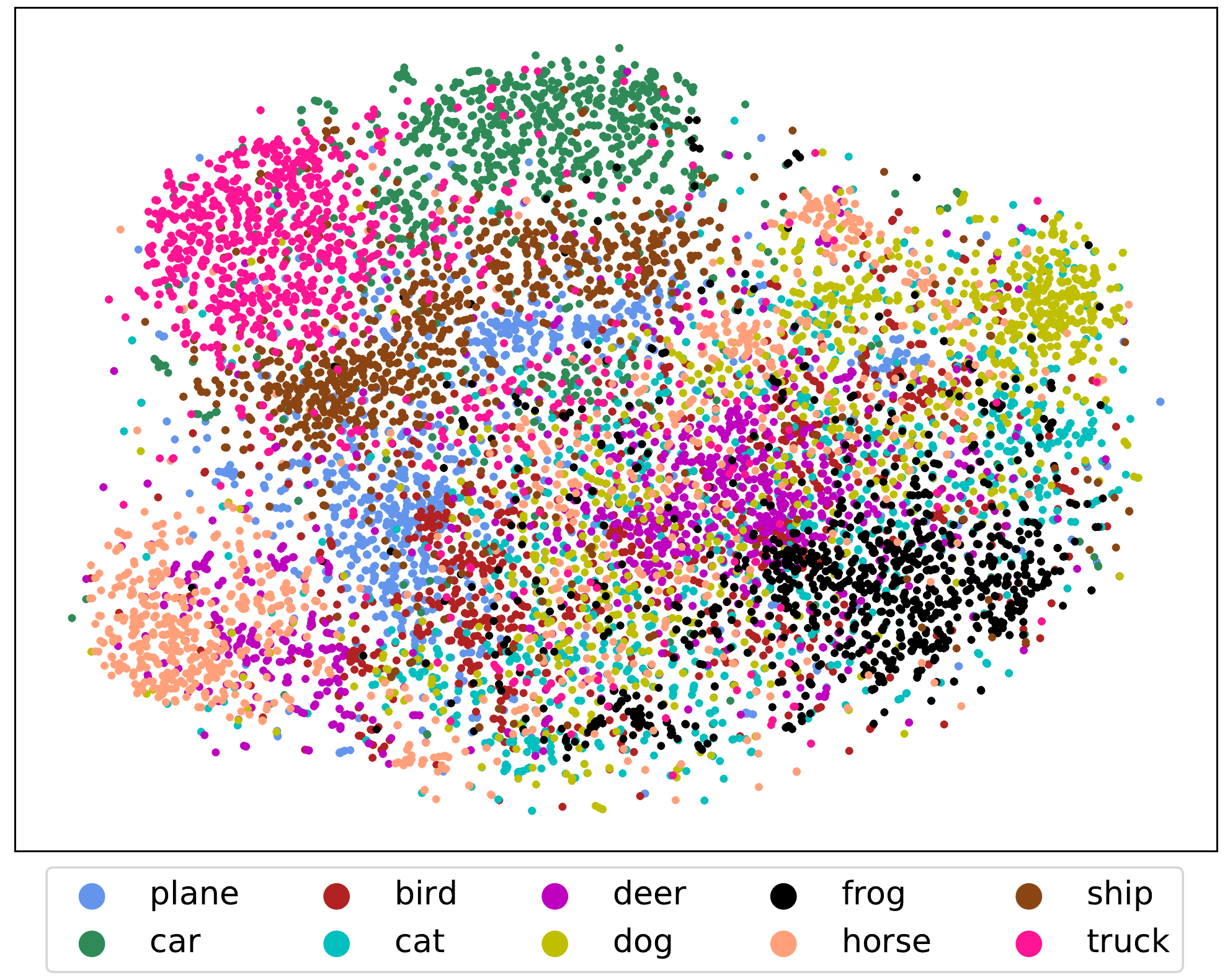}}
  \subfigure[Agg. Online, Update Online ]{\label{fig:online-online}\includegraphics[width=37mm]{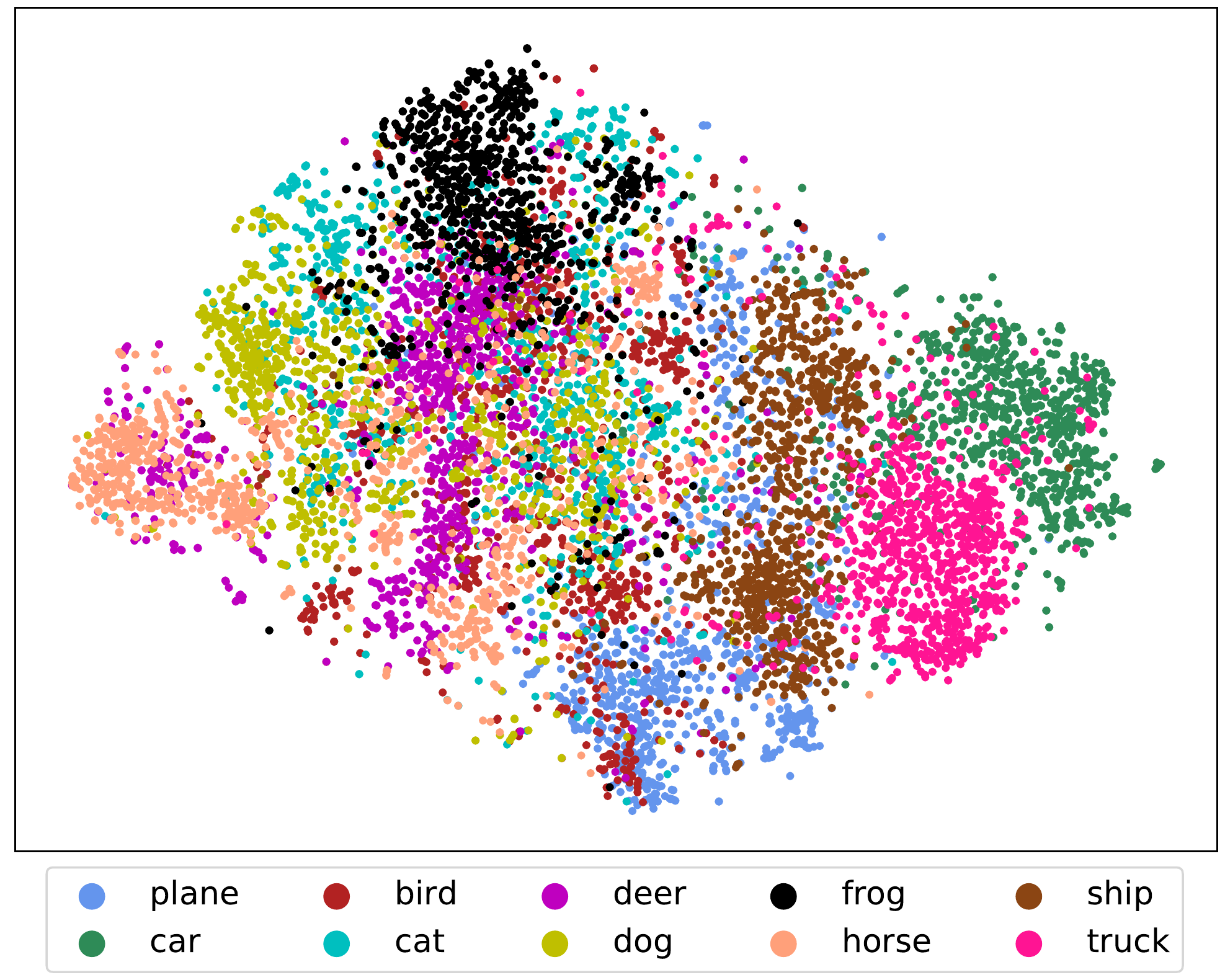}}
  \subfigure[FedU]{\label{fig:tsne-fedu}\includegraphics[width=37mm]{charts/tsne_whole_0.2.pdf}}
\end{center}
  \caption{T-SNE visualization of representations learned from four methods: (a) Aggregate the online encoder and update the target encoder; (b) Aggregate the target encoder and update the online encoder; (c) Aggregate and update the online encoder; (d) Our proposed FedU. (a), (b), and (c) always use the local predictor, while (d) uses DAPU to dynamically update the predictor. Aggregating and updating with the online encoder (c) achieves better clustering results than (a) and (b). FedU with DAPU (d) further improves the results.}
  \label{fig:tsne-ablation}
\end{figure*}

\subsection{Semi-Supervised Learning}

We evaluate the representation learned from FedU on the semi-supervised protocol described in \cite{zhai2019s4l, chen2020simclr}, targeting the federated scenarios that only a small subset of data are labeled. We consider two semi-supervised learning settings: 1\% or 10\% are labeled. We first obtain the representations using FedU and other methods without the labeled data. Then, instead of fixing the model, we fine-tune the whole model with an additional new classifier in the semi-supervised protocol using the labeled data for 100 epochs. The methods compared are similar to the ones defined in the linear evaluation section. 



As reported in Table \ref{tab:semi-sup}, FedU outperforms other methods except centralized training on semi-supervised evaluation protocol. Supervised federated learning with FedAvg using only 1\% or 10\% of data has poor performance. On evaluation using the CIFAR-10 dataset, it outperforms FedCA \cite{zhang2020fedca} by more than 22\% on IID data and more than 40\% on non-IID data. Besides, it consistently outperforms single client training by around 3\% regardless of the settings.

\subsection{Transfer Learning}
\label{sec:transfer-learning}

We assess the generalizability of representations learned from FedU by evaluating the learned representations on different classification datasets. Specifically, we learn representations on the Mini-ImageNet \cite{vinyals2016mini-imagenet} dataset and evaluate how well they can be transferred to CIFAR datasets. After training, we fine-tune representations on the target datasets for 100 epochs. 

Table \ref{tab:transfer-learning} compares the transfer learning results using ResNet-50 \cite{he2016resnet}. Even though FedU only slightly outperforms FedCA on the CIFAR-10 dataset because even the random initialization can achieve relatively good performance, it achieves much better performance on the CIFAR-100 dataset, the closest to centralized unsupervised representation learning.

\begingroup
\setlength{\tabcolsep}{0.22em}
\begin{table}[t]
\begin{center}
\begin{tabular}{lccccc}
\hline
\multicolumn{1}{l}{\multirow{2}{*}{Method}} &
\multicolumn{2}{c}{CIFAR-10} &
\multicolumn{1}{c}{} &
\multicolumn{2}{c}{CIFAR-100}
\\ 
\cline{2-3} \cline{5-6}
\multicolumn{1}{c}{} & IID & Non-IID & & IID & Non-IID
\\
\hline 
Random Init. & 93.79 & - & & 70.52 & - \\
Single client training & 94.67 & 94.33 & & 75.25 & 75.14 \\
FedSimCLR \cite{chen2020simclr} \cite{zhang2020fedca} & 94.87 & 93.97 & & 71.85 & 70.91 \\ 
FedCA \cite{zhang2020fedca} & 94.94 & 94.16 & & 71.98 & 71.32 \\ 
FedU (ours) & \textbf{95.00} & \textbf{94.83} & & \textbf{75.57} & \textbf{75.60} \\ 
\hline
BYOL \cite{grill2020byol} (Centralized) & 95.37 & - & & 77.02 & - \\
\hline
\end{tabular}
\end{center}
\caption{Top-1 accuracy (\%) comparison under the transfer learning protocol: transferring from Mini-ImageNet to CIFAR datasets using ResNet-50. FedU outperforms other baseline methods.}
\label{tab:transfer-learning}
\end{table}    
\endgroup





\section{Ablation Study}

In this section, we conduct ablation studies on the communication protocol, divergence-aware predictor update (DAPU), and hyperparameters of FedU. These ablations give intuitions of FedU's behavior and performance.


\subsection{Online Encoder vs. Target Encoder}
\label{sec:ablation-encoders}

\begin{table}[t]
\begin{center}
\begin{tabular}{llcc}
\hline

\multicolumn{1}{l}{\multirow{2}{*}{Aggregate}} &
\multicolumn{1}{l}{\multirow{2}{*}{Update}} &

\multicolumn{2}{c}{Accuracy (\%)}
\\ 
\cline{3-4} 
\multicolumn{1}{c}{} &  & Global Pred. & Local Pred.

\\
\hline 

\textbf{Online} & \textbf{Online} & \textbf{84.07} & \textbf{82.18} \\ 
Online & Target & 9.99 & 19.22 \\ 
Online & Both & 81.24 & 18.23 \\ 
Target & Online & 82.10 & 78.06 \\ 
Target & Target & 9.99 & 25.02 \\ 
Target & Both & 82.32 & 29.03 \\ 

\hline
\end{tabular}
\end{center}
\caption{Top-1 accuracy comparison of using the \textit{online encoder} or \textit{target encoder} for aggregation and update. Both means updating both encoders. The predictors are updated with either the global or local predictor. Aggregating and updating the online encoder achieves the best performance.}
\label{tab:abs-encoders}
\end{table}    







In Section \ref{sec:encoder}, FedU uploads the online encoders for aggregation and updates clients' online encoders with the aggregated global encoder. To empirically verify this hypothesis, we conduct training with twelve combinations of the encoder to upload (online or target encoder), the encoder to update (online, target, or both), and the choice of the predictor (local or global predictor) using ResNet-50 on the CIFAR-10 non-IID setting. Local/global predictor means that clients always update the predictor with parameters of the local/global predictor.

As shown in Table \ref{tab:abs-encoders}, aggregating and updating the online encoder achieves the best performance, regardless of the choice of the predictor. Besides, we compare the t-SNE visualization of representations in Figure \ref{fig:tsne-ablation}. Aggregating and updating the online encoder (Figure \ref{fig:online-online}) achieves better clustering results than Figure \ref{fig:online-target} and Figure \ref{fig:target-online}. These results further verify our hypothesis on the behaviors and intuitions of encoders in clients. The target encoder provides regression targets for the online encoder, so only updating it results in poor performance. Updating both the online and target encoders achieves competitive results, but it is not comparable to the best performance. 


\subsection{Divergence-aware Predictor Update}



\begin{figure}[t]
\begin{center}
  \subfigure{\label{fig:dapu}\includegraphics[width=40mm]{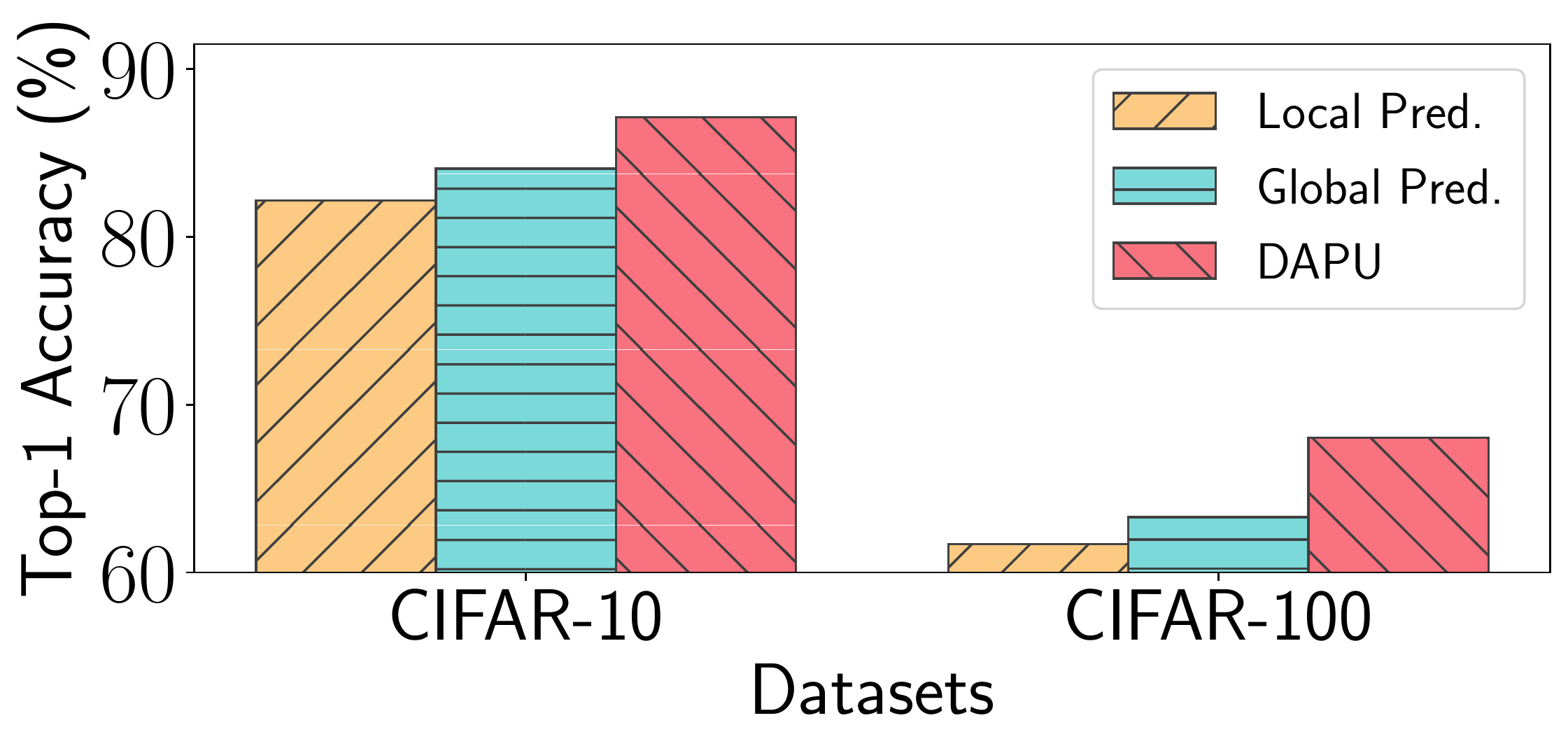}}
  \subfigure{\label{fig:threshold}\includegraphics[width=40mm]{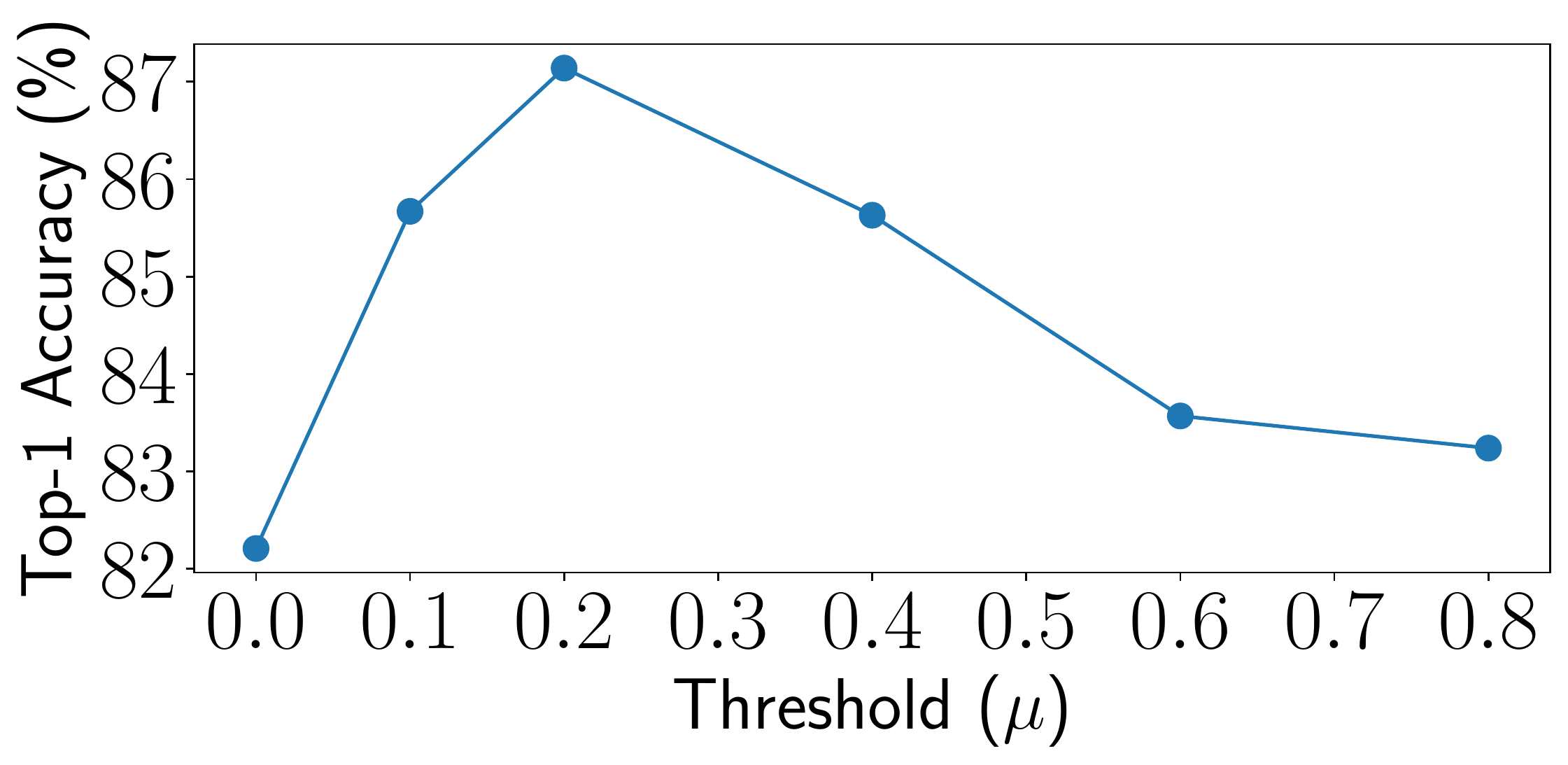}}
\end{center}
  \caption{Ablation study on divergence-aware predictor update (DAPU): (a) Compare DAPU with always updating with the local or global predictor; (b) Analysis of threshold $\mu$.}
  \label{fig:abs-dapi}
\end{figure}


FedU contains a new module, divergence-aware predictor update (DAPU), to dynamically update clients' predictors with either their local predictors $p_\theta$ or the global predictor $p_\phi$. To understand the effectiveness of DAPU for non-IID data, we compare DAPU with two static update methods: always updating with the local predictor $p_\theta$ and always updating with the global predictor $p_\phi$. Besides, we also evaluate the impact of threshold $\mu$. By default, we aggregate and update the online encoders.

\textbf{Compare with Other Model Update Methods} Figure \ref{fig:dapu} demonstrates that DAPU outperforms the other two model update methods by around 5\%, under non-IID setting on CIFAR datasets. Moreover, the t-SNE visualization of representations of FedU with DAPU (Figure \ref{fig:tsne-fedu}) is better than always updating with the local predictor (Figure \ref{fig:online-online}). These results demonstrate that our proposed DAPU is effective and significant. Next, we present the ablations on the values of threshold $\mu$.





\textbf{Impact of Threshold} Figure \ref{fig:threshold} compares the impact of values of $\mu$ on the CIFAR-10 dataset. FedU reaches the best performance when $\mu = 0.2$. As discussed in Section \ref{sec:encoder}, the global encoder is the average of online encoders and the online encoders are updated with it in the next round. The divergence between the global encoder $\phi^{r-1}$ and online encoder $\theta^r$ reduces as training proceeds because the encoders obtain higher generalizability and converge gradually. Hence, an optimal threshold $\mu$ exists to balance updating with the local predictor $p_\theta$ or the global predictor $p_\phi$. On the one hand, large $\mu$ results in updating with the global predictor too early when the divergence is still significant. On the other hand, small $\mu$ results in updating with the global predictor too late. Despite that the impact of divergence varies depending on datasets, these results also hold for the CIFAR-100 dataset (provided in the supplementary).


\subsection{Analysis of FedU}

\begin{figure}[t]
\begin{center}
  \subfigure[Impact of Local Epochs]{\label{fig:local-epoch}\includegraphics[width=39mm]{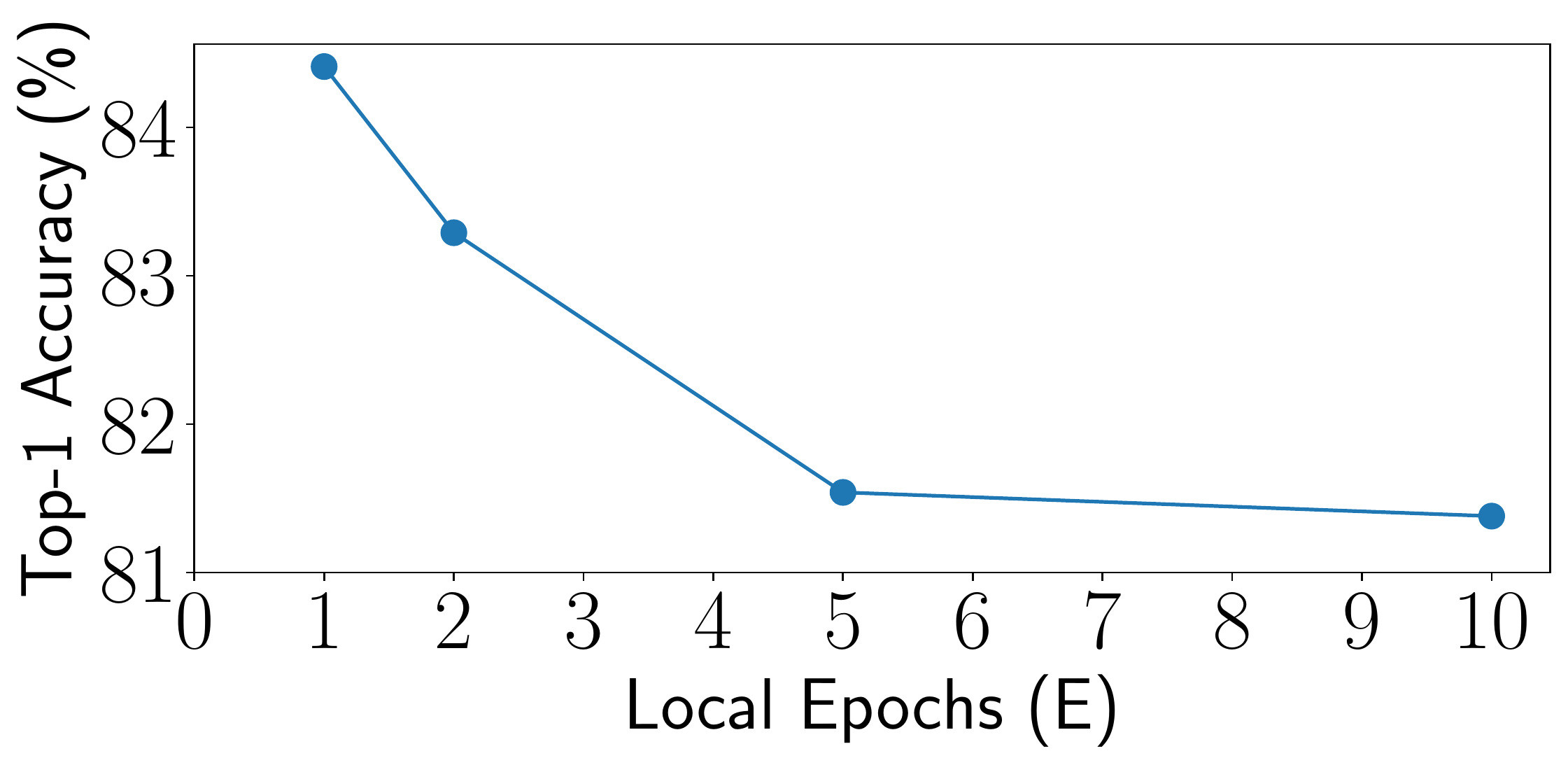}}
  \subfigure[Impact of Total Rounds]{\label{fig:total-round}\includegraphics[width=41mm]{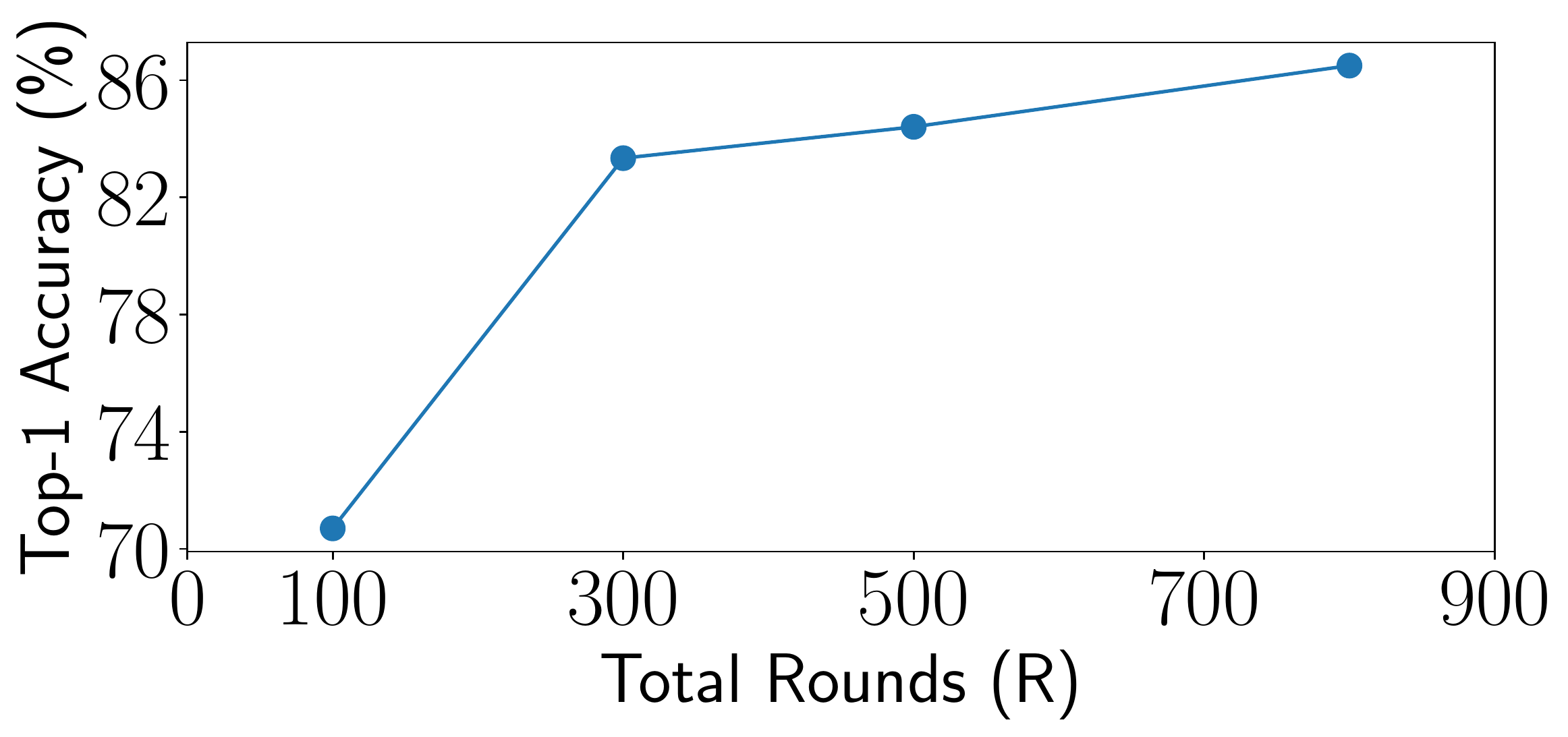}}
\end{center}
  \caption{Analysis of (a) impact of local epochs and (b) impact of total training rounds on the performance of FedU.}
  \label{fig:ablation-fedu}
\end{figure}

\textbf{Impact of Local Epoch} Local epochs $E$ represents the trade-off between accuracy and communication-cost. We measure the impact of different numbers of $E$ by fixing the total computation executed in each client to be 500 epochs. As such, for each training round, each client trains $E$ epochs locally and then transmits the model updates to the server. The total communication rounds are $\frac{500}{E}$. Figure \ref{fig:local-epoch} illustrates that the accuracy decreases as $E$ increases. Smaller local epoch $E$ achieves better performance but causes higher communication costs. In scenarios that network bandwidth is not bounded, we suggest using $E = 1$ as the default configuration for better performance.

\textbf{Impact of Training Rounds} We investigate the impact of training rounds by fixing $E = 1$ and varies total training rounds $R$ from 100 to 800. Figure \ref{fig:total-round} shows that increasing total training rounds results in better performance. The accuracy improvement is especially significant when the number of total training rounds is small. FedU outperforms other methods with $R = 100$ in Section \ref{sec:experiments}, though the accuracy can be further improved by increasing $R$.

\section{Conclusion}

In this work, we introduce a new framework, FedU, to learn generic representations from multiple parties by leveraging decentralized unlabeled data. It includes two simple but effective methods to tackle the non-IID challenge of decentralized data. Firstly, we design the communication protocol to aggregate and update the online encoders. Secondly, we propose a novel module, divergence-aware predictor update (DAPU), to dynamically decide how to update the predictors. We extensively evaluate FedU and conducts ablation studies to illustrate the intuitions and behaviors of the framework. FedU outperforms existing methods on all evaluation protocols. For future work, we will consider applying FedU on larger scale datasets and evaluating specific application scenarios. We hope that FedU will encourage the community to explore learning visual representations from decentralized data under privacy constraints. 

\noindent\textbf{Acknowledgements} This study is supported by 1) the RIE2020 Industry Alignment Fund – Industry Collaboration Projects (IAF-ICP) Funding Initiative, as well as cash and in-kind contribution from the industry partner(s); 2) the National Research Foundation, Singapore, and the Energy Market Authority, under its Energy Programme (EP Award $<$NRF2017EWT-EP003-023$>$); 3) Singapore MOE under its Tier 1 grant call, Reference number RG96/20.

{\small
\bibliographystyle{ieee_fullname}
\bibliography{egbib}

\begin{thebibliography}{10}\itemsep=-1pt

\bibitem{chen2020simclr}
Ting Chen, Simon Kornblith, Mohammad Norouzi, and Geoffrey Hinton.
\newblock A simple framework for contrastive learning of visual
  representations.
\newblock In {\em International conference on machine learning}, pages
  1597--1607. PMLR, 2020.

\bibitem{chen2020simsiam}
Xinlei Chen and Kaiming He.
\newblock Exploring simple siamese representation learning.
\newblock {\em arXiv preprint arXiv:2011.10566}, 2020.

\bibitem{gdpr}
Bart Custers, Alan~M. Sears, Francien Dechesne, Ilina Georgieva, Tommaso Tani,
  and Simone van~der Hof.
\newblock {\em EU Personal Data Protection in Policy and Practice}.
\newblock Springer, 2019.

\bibitem{imagenet_cvpr09}
J. Deng, W. Dong, R. Socher, L.-J. Li, K. Li, and L. Fei-Fei.
\newblock {ImageNet: A Large-Scale Hierarchical Image Database}.
\newblock In {\em CVPR09}, 2009.

\bibitem{doersch2015unsup-context-prediction}
Carl Doersch, Abhinav Gupta, and Alexei~A Efros.
\newblock Unsupervised visual representation learning by context prediction.
\newblock In {\em Proceedings of the IEEE international conference on computer
  vision}, pages 1422--1430, 2015.

\bibitem{gidaris2018unsup-image-rotation}
Spyros Gidaris, Praveer Singh, and Nikos Komodakis.
\newblock Unsupervised representation learning by predicting image rotations.
\newblock {\em arXiv preprint arXiv:1803.07728}, 2018.

\bibitem{NIPS2014GAN}
Ian Goodfellow, Jean Pouget-Abadie, Mehdi Mirza, Bing Xu, David Warde-Farley,
  Sherjil Ozair, Aaron Courville, and Yoshua Bengio.
\newblock Generative adversarial nets.
\newblock In Z. Ghahramani, M. Welling, C. Cortes, N. Lawrence, and K.~Q.
  Weinberger, editors, {\em Advances in Neural Information Processing Systems},
  volume~27. Curran Associates, Inc., 2014.

\bibitem{grill2020byol}
Jean-Bastien Grill, Florian Strub, Florent Altch\'{e}, Corentin Tallec, Pierre
  Richemond, Elena Buchatskaya, Carl Doersch, Bernardo Avila~Pires, Zhaohan
  Guo, Mohammad Gheshlaghi~Azar, Bilal Piot, koray kavukcuoglu, Remi Munos, and
  Michal Valko.
\newblock Bootstrap your own latent - a new approach to self-supervised
  learning.
\newblock In H. Larochelle, M. Ranzato, R. Hadsell, M.~F. Balcan, and H. Lin,
  editors, {\em Advances in Neural Information Processing Systems}, volume~33,
  pages 21271--21284. Curran Associates, Inc., 2020.

\bibitem{hadsell2006constrastive-loss}
Raia Hadsell, Sumit Chopra, and Yann LeCun.
\newblock Dimensionality reduction by learning an invariant mapping.
\newblock In {\em 2006 IEEE Computer Society Conference on Computer Vision and
  Pattern Recognition (CVPR'06)}, volume~2, pages 1735--1742. IEEE, 2006.

\bibitem{he2020moco}
Kaiming He, Haoqi Fan, Yuxin Wu, Saining Xie, and Ross Girshick.
\newblock Momentum contrast for unsupervised visual representation learning.
\newblock In {\em Proceedings of the IEEE/CVF Conference on Computer Vision and
  Pattern Recognition}, pages 9729--9738, 2020.

\bibitem{he2016resnet}
Kaiming He, Xiangyu Zhang, Shaoqing Ren, and Jian Sun.
\newblock Deep residual learning for image recognition.
\newblock In {\em Proceedings of the IEEE conference on computer vision and
  pattern recognition}, pages 770--778, 2016.

\bibitem{jin2020fed-unlabeled-survey}
Yilun Jin, Xiguang Wei, Yang Liu, and Qiang Yang.
\newblock Towards utilizing unlabeled data in federated learning: A survey and
  prospective.
\newblock {\em arXiv e-prints}, pages arXiv--2002, 2020.

\bibitem{kingma2013auto-encoding}
Diederik~P Kingma and Max Welling.
\newblock Auto-encoding variational bayes.
\newblock {\em arXiv preprint arXiv:1312.6114}, 2013.

\bibitem{kolesnikov2019revisiting}
Alexander Kolesnikov, Xiaohua Zhai, and Lucas Beyer.
\newblock Revisiting self-supervised visual representation learning.
\newblock In {\em Proceedings of the IEEE/CVF Conference on Computer Vision and
  Pattern Recognition}, pages 1920--1929, 2019.

\bibitem{cifar10-2009}
Alex Krizhevsky, Geoffrey Hinton, et~al.
\newblock Learning multiple layers of features from tiny images.
\newblock 2009.

\bibitem{Li2020FedChallenges}
Tian Li, Anit~Kumar Sahu, Ameet Talwalkar, and Virginia Smith.
\newblock Federated learning: Challenges, methods, and future directions.
\newblock {\em IEEE Signal Processing Magazine}, 37:50--60, 2020.

\bibitem{fedprox}
Tian Li, Anit~Kumar Sahu, Manzil Zaheer, Maziar Sanjabi, Ameet Talwalkar, and
  Virginia Smith.
\newblock Federated optimization in heterogeneous networks.
\newblock {\em Proceedings of Machine Learning and Systems}, 2:429--450, 2020.

\bibitem{li2019brain-tumor1}
Wenqi Li, Fausto Milletar{\`\i}, Daguang Xu, Nicola Rieke, Jonny Hancox, Wentao
  Zhu, Maximilian Baust, Yan Cheng, S{\'e}bastien Ourselin, M~Jorge Cardoso,
  et~al.
\newblock Privacy-preserving federated brain tumour segmentation.
\newblock In {\em International Workshop on Machine Learning in Medical
  Imaging}, pages 133--141. Springer, 2019.

\bibitem{long2015segmentation}
Jonathan Long, Evan Shelhamer, and Trevor Darrell.
\newblock Fully convolutional networks for semantic segmentation.
\newblock In {\em Proceedings of the IEEE conference on computer vision and
  pattern recognition}, pages 3431--3440, 2015.

\bibitem{luo2019real-obj-detect}
Jiahuan Luo, Xueyang Wu, Yun Luo, Anbu Huang, Yunfeng Huang, Yang Liu, and
  Qiang Yang.
\newblock Real-world image datasets for federated learning.
\newblock {\em arXiv preprint arXiv:1910.11089}, 2019.

\bibitem{fedavg}
Brendan McMahan, Eider Moore, Daniel Ramage, Seth Hampson, and
  Blaise~Ag{\"{u}}era y Arcas.
\newblock Communication-efficient learning of deep networks from decentralized
  data.
\newblock In Aarti Singh and Xiaojin~(Jerry) Zhu, editors, {\em Proceedings of
  the 20th International Conference on Artificial Intelligence and Statistics,
  {AISTATS} 2017, 20-22 April 2017, Fort Lauderdale, FL, {USA}}, volume~54 of
  {\em Proceedings of Machine Learning Research}, pages 1273--1282. {PMLR},
  2017.

\bibitem{noroozi2016jigsaw}
Mehdi Noroozi and Paolo Favaro.
\newblock Unsupervised learning of visual representations by solving jigsaw
  puzzles.
\newblock In {\em European conference on computer vision}, pages 69--84.
  Springer, 2016.

\bibitem{oord2018contrastive-predictive-coding}
Aaron van~den Oord, Yazhe Li, and Oriol Vinyals.
\newblock Representation learning with contrastive predictive coding.
\newblock {\em arXiv preprint arXiv:1807.03748}, 2018.

\bibitem{oquab2014transferring-downstream}
Maxime Oquab, Leon Bottou, Ivan Laptev, and Josef Sivic.
\newblock Learning and transferring mid-level image representations using
  convolutional neural networks.
\newblock In {\em Proceedings of the IEEE conference on computer vision and
  pattern recognition}, pages 1717--1724, 2014.

\bibitem{paszke2017pytorch}
Adam Paszke, Sam Gross, Soumith Chintala, Gregory Chanan, Edward Yang, Zachary
  DeVito, Zeming Lin, Alban Desmaison, Luca Antiga, and Adam Lerer.
\newblock Automatic differentiation in pytorch.
\newblock 2017.

\bibitem{pathak2016inpainting}
Deepak Pathak, Philipp Krahenbuhl, Jeff Donahue, Trevor Darrell, and Alexei~A
  Efros.
\newblock Context encoders: Feature learning by inpainting.
\newblock In {\em Proceedings of the IEEE conference on computer vision and
  pattern recognition}, pages 2536--2544, 2016.

\bibitem{radford2015adversarial}
Alec Radford, Luke Metz, and Soumith Chintala.
\newblock Unsupervised representation learning with deep convolutional
  generative adversarial networks.
\newblock {\em arXiv preprint arXiv:1511.06434}, 2015.

\bibitem{sheller2018brain-tumor2}
Micah~J Sheller, G~Anthony Reina, Brandon Edwards, Jason Martin, and Spyridon
  Bakas.
\newblock Multi-institutional deep learning modeling without sharing patient
  data: A feasibility study on brain tumor segmentation.
\newblock In {\em International MICCAI Brainlesion Workshop}, pages 92--104.
  Springer, 2018.

\bibitem{simonyan2014very-deep-cnn}
Karen Simonyan and Andrew Zisserman.
\newblock Very deep convolutional networks for large-scale image recognition.
\newblock {\em arXiv preprint arXiv:1409.1556}, 2014.

\bibitem{van2020fedae}
Bram van Berlo, Aaqib Saeed, and Tanir Ozcelebi.
\newblock Towards federated unsupervised representation learning.
\newblock In {\em Proceedings of the Third ACM International Workshop on Edge
  Systems, Analytics and Networking}, pages 31--36, 2020.

\bibitem{vincent2008autoencoder}
Pascal Vincent, Hugo Larochelle, Yoshua Bengio, and Pierre-Antoine Manzagol.
\newblock Extracting and composing robust features with denoising autoencoders.
\newblock In {\em Proceedings of the 25th international conference on Machine
  learning}, pages 1096--1103, 2008.

\bibitem{vinyals2016mini-imagenet}
Oriol Vinyals, Charles Blundell, Timothy Lillicrap, Koray Kavukcuoglu, and Daan
  Wierstra.
\newblock Matching networks for one shot learning.
\newblock {\em arXiv preprint arXiv:1606.04080}, 2016.

\bibitem{wu2018unsupervised-instance}
Zhirong Wu, Yuanjun Xiong, Stella~X Yu, and Dahua Lin.
\newblock Unsupervised feature learning via non-parametric instance
  discrimination.
\newblock In {\em Proceedings of the IEEE Conference on Computer Vision and
  Pattern Recognition}, pages 3733--3742, 2018.

\bibitem{zeiler2014visualizing-cnn}
Matthew~D Zeiler and Rob Fergus.
\newblock Visualizing and understanding convolutional networks.
\newblock In {\em European conference on computer vision}, pages 818--833.
  Springer, 2014.

\bibitem{zhai2019s4l}
Xiaohua Zhai, Avital Oliver, Alexander Kolesnikov, and Lucas Beyer.
\newblock S4l: Self-supervised semi-supervised learning.
\newblock In {\em Proceedings of the IEEE/CVF International Conference on
  Computer Vision}, pages 1476--1485, 2019.

\bibitem{zhang2020fedca}
Fengda Zhang, Kun Kuang, Zhaoyang You, Tao Shen, Jun Xiao, Yin Zhang, Chao Wu,
  Yueting Zhuang, and Xiaolin Li.
\newblock Federated unsupervised representation learning.
\newblock {\em arXiv preprint arXiv:2010.08982}, 2020.

\bibitem{zhang2016colorful-image}
Richard Zhang, Phillip Isola, and Alexei~A Efros.
\newblock Colorful image colorization.
\newblock In {\em European conference on computer vision}, pages 649--666.
  Springer, 2016.

\bibitem{zhao2018non-iid}
Yue Zhao, Meng Li, Liangzhen Lai, Naveen Suda, Damon Civin, and Vikas Chandra.
\newblock Federated learning with non-iid data.
\newblock {\em CoRR}, abs/1806.00582, 2018.

\bibitem{zhuang2021easyfl}
Weiming Zhuang, Xin Gan, Yonggang Wen, and Shuai Zhang.
\newblock Easyfl: A low-code federated learning platform for dummies.
\newblock {\em arXiv preprint arXiv:2105.07603}, 2021.

\bibitem{zhuang2021fedfr}
Weiming ZHuang, Xin Gan, Yonggang Wen, and Shuai Zhang.
\newblock Towards unsupervised domain adaptation for deep face recognition
  under privacy constraints via federated learning.
\newblock In {\em Proceedings of the IEEE/CVF Conference on Computer Vision and
  Pattern Recognition}, 2021.

\bibitem{zhuang2021fedureid}
Weiming Zhuang, Yonggang Wen, and Shuai Zhang.
\newblock Joint optimization in edge-cloud continuum for federated unsupervised
  person re-identification.
\newblock In {\em Proceedings of the 29th ACM International Conference on
  Multimedia}, 2021.

\bibitem{zhuang2020fedreid}
Weiming Zhuang, Yonggang Wen, Xuesen Zhang, Xin Gan, Daiying Yin, Dongzhan
  Zhou, Shuai Zhang, and Shuai Yi.
\newblock Performance optimization of federated person re-identification via
  benchmark analysis.
\newblock In {\em Proceedings of the 28th ACM International Conference on
  Multimedia}, pages 955--963, 2020.

\end{thebibliography}
}

\end{document}


\title{Supplementary for \textit{Collaborative Unsupervised Visual Representation Learning from Decentralized Data}}

\author{Weiming Zhuang$^{1, 3}$ \quad Xin Gan$^{2}$ \quad Yonggang Wen$^{2}$ \quad Shuai Zhang$^{3}$ \quad Shuai Yi$^{3}$\\
$^{1}$S-Lab, Nanyang Technological University, $^{2}$Nanyang Technological University, 
$^{3}$SenseTime Research \\
{\tt\small \{weiming001,ganx0005\}@e.ntu.edu.sg,ygwen@ntu.edu.sg,\{zhangshuai,yishuai\}@sensetime.com}
}

\maketitle
\ificcvfinal\thispagestyle{empty}\fi

\begin{abstract}

This supplementary material provides more implementation details and presents more experimental results, including ablation studies on DAPU, batch size, and the number of clients. We also present the convergence of FedU and more t-SNE visualization of representations of different settings.

\end{abstract}

\section{Implementation Details}

In this section, we provide more implementation details on image augmentations, semi-supervised evaluation, and transfer learning evaluation.

\textbf{Image Augmentations} We adopt the image augmentations from BYOL \cite{grill2020byol} and SimCLR \cite{chen2020simclr}. We select a random patch of the image and resize it to 32x32 for CIFAR datasets. After that, two transformations are applied to the image: a random horizontal flip and a color distortion. 

\textbf{Semi-supervised Learning} In semi-supervised evaluation protocol, we train models with only unlabeled data --- excluding the 1\% or 10\% labeled data. These 1\% and 10\% labeled data are only used in fine-tuning the trained models with an additional classifier.

\textbf{Transfer Learning} In transfer learning evaluation protocol, we train models using Mini-ImageNet \cite{vinyals2016mini-imagenet} dataset and fine-tune on CIFAR \cite{cifar10-2009} datasets. Images in Mini-ImageNet have size 84x84, while images in CIFAR datasets are 32x32. We scale image size of CIFAR datasets to be 84x84 in fine-tuning.

\section{Experimental Results}

\textbf{Communication Protocol} Table \ref{tab:abs-encoders-cifar100} shows that aggregating and uploading the online encoder achieves the best performance. We run these experiments with ResNet-18 on CIFAR-100 non-IID setting. It complements the results of Table 4 in the main manuscript.

\begin{table}[t]
  \begin{center}
  \begin{tabular}{llcc}
  \hline
  \multicolumn{1}{l}{Aggregate} &
  \multicolumn{1}{l}{Update} &
  \multicolumn{1}{c}{Accuracy (\%)}
  \\
  \hline 
  \textbf{Online} & \textbf{Online} & \textbf{62.86} \\ 
  Online & Target & 3.44 \\ 
  Online & Both & 60.78 \\ 
  Target & Online & 48.10 \\ 
  Target & Target & 10.27 \\ 
  Target & Both & 61.41 \\ 
  \hline
  \end{tabular}
  \end{center}
  \caption{Top-1 accuracy comparison of using the \textit{online encoder} or \textit{target encoder} for aggregation and update. Both means updating both encoders. Aggregating and updating the online encoder achieves the best performance.}
  \label{tab:abs-encoders-cifar100}
\end{table}    


\textbf{Divergence-aware Predictor Update} Table \ref{tab:abs-predictor} shows that DAPU outperforms always updating the predictors of clients using either the local or global predictor on both IID and non-IID settings. It complements the results of Figure 5(a) in the main manuscript, which only presents that DAPU outperforms other methods on the non-IID setting. Besides, Figure \ref{fig:threshold-cifar100} presents the impact of threshold $\mu$ on the non-IID setting of the CIFAR-100 dataset, complementing the results on the CIFAR-10 dataset in the main manuscript (Figure 5(b)). It also indicates that $\mu = 0.2$ achieves the best performance. We run these experiments with $E = 1$ and $R = 800$.

\begingroup
\setlength{\tabcolsep}{0.5em}
\begin{table}[t]
\begin{center}
\begin{tabular}{lccccc}
\hline
\multicolumn{1}{l}{\multirow{2}{*}{Update Method}} &
\multicolumn{2}{c}{CIFAR-10} &
\multicolumn{1}{c}{} &
\multicolumn{2}{c}{CIFAR-100}
\\ 
\cline{2-3} \cline{5-6} 
\multicolumn{1}{c}{} & Non-IID & IID & & Non-IID & IID
\\
\hline 

Local Pred. & 82.18 & 91.29 & & 61.69 & 67.41 \\ 
Global Pred. & 84.07 & 91.41 & & 63.30 & 67.56 \\ 
DAPU & \textbf{87.14} & \textbf{93.13} & & \textbf{68.02} & \textbf{67.66} \\ 

\hline
\end{tabular}
\end{center}
\caption{Top-1 accuracy comparison of DAPU and always updating with the local or global predictor. DAPU outperforms other methods in all settings.}
\label{tab:abs-predictor}
\end{table} 
\endgroup

\begin{figure*}[t]
  \begin{center}
    \subfigure[Agg. Online, Update Target]{\label{fig:online-target-global}\includegraphics[width=40mm]{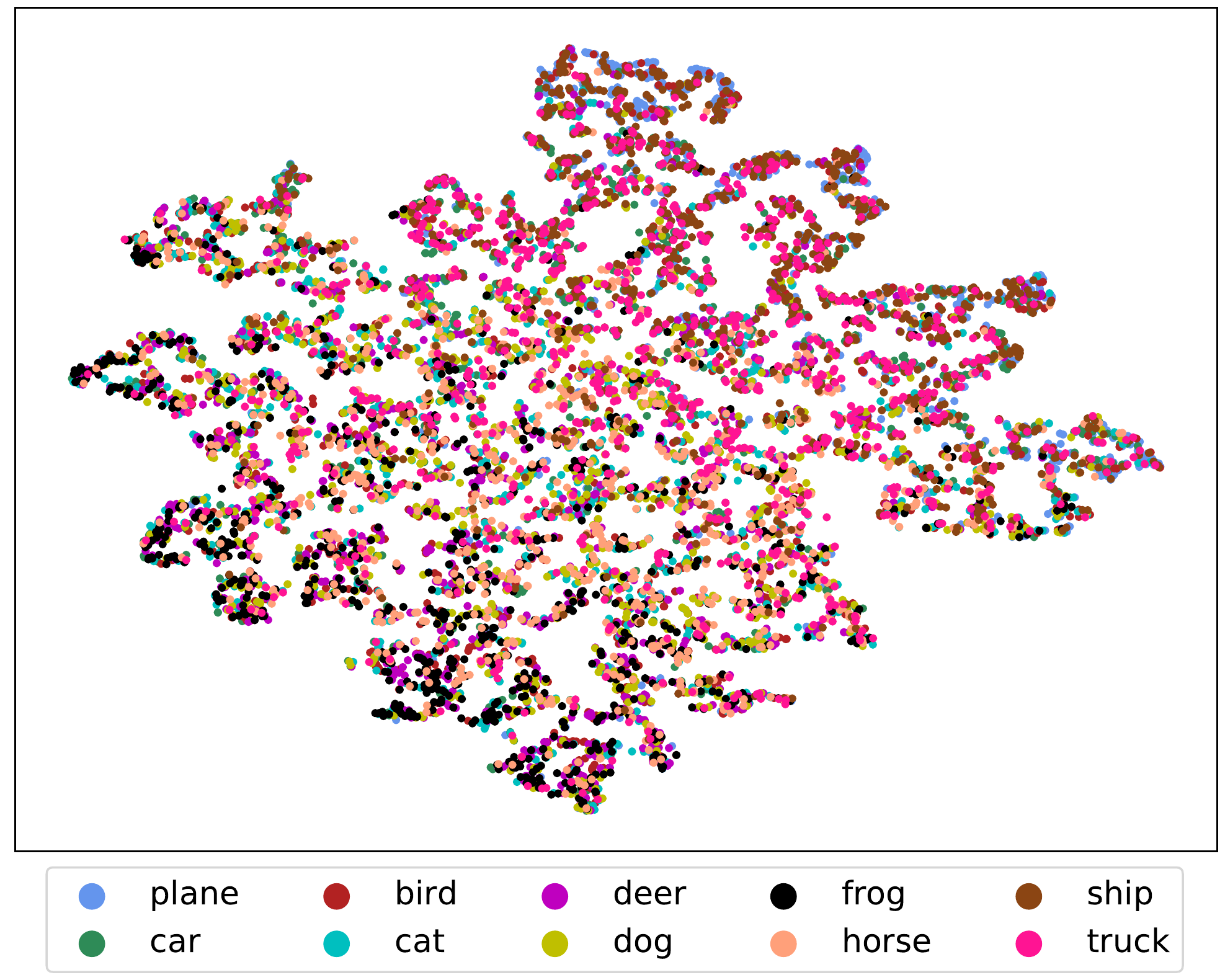}}
    \subfigure[Agg. Target, Update Online]{\label{fig:target-online-global}\includegraphics[width=40mm]{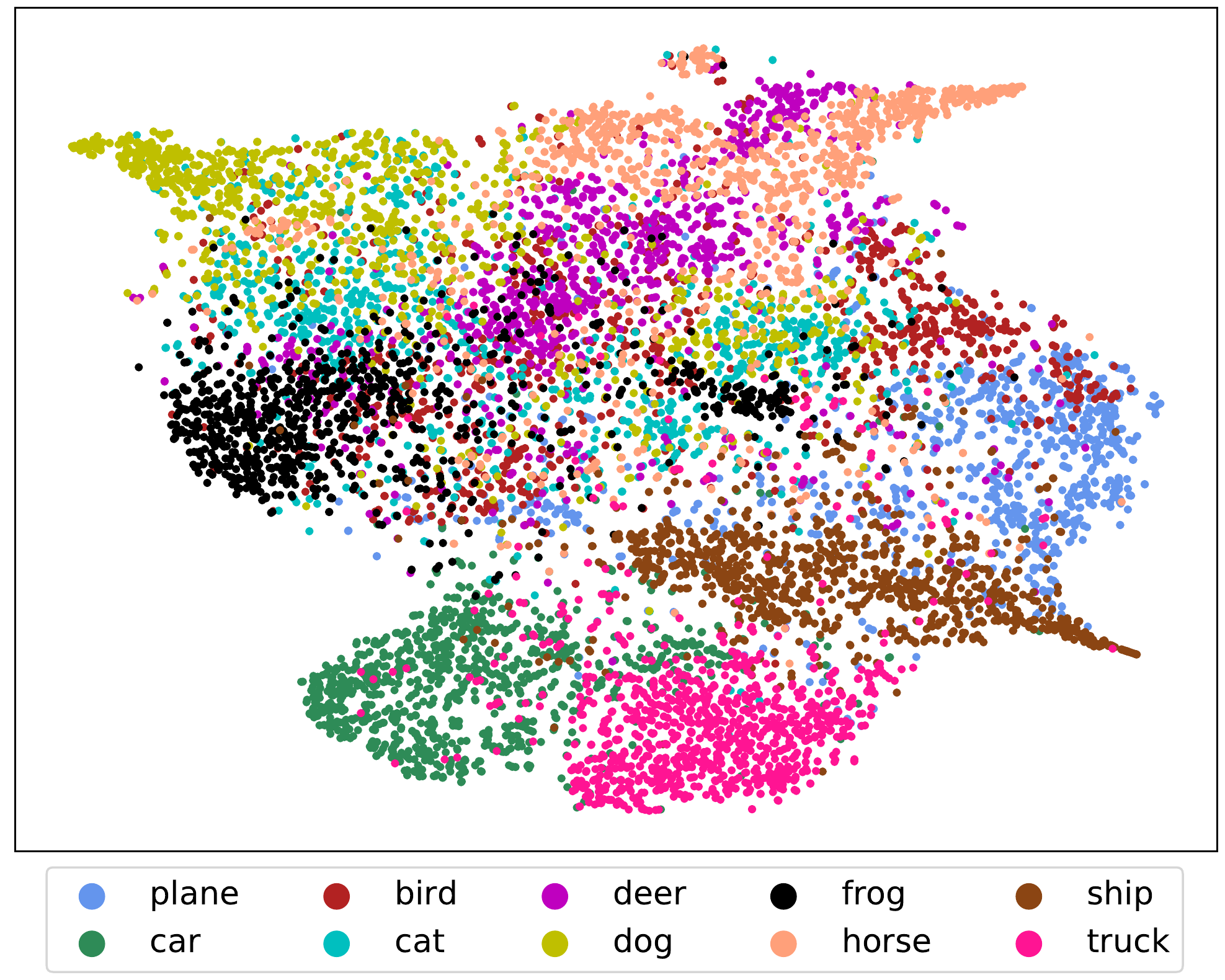}}
    \subfigure[Agg. Online, Update Target]{\label{fig:online-online-global}\includegraphics[width=40mm]{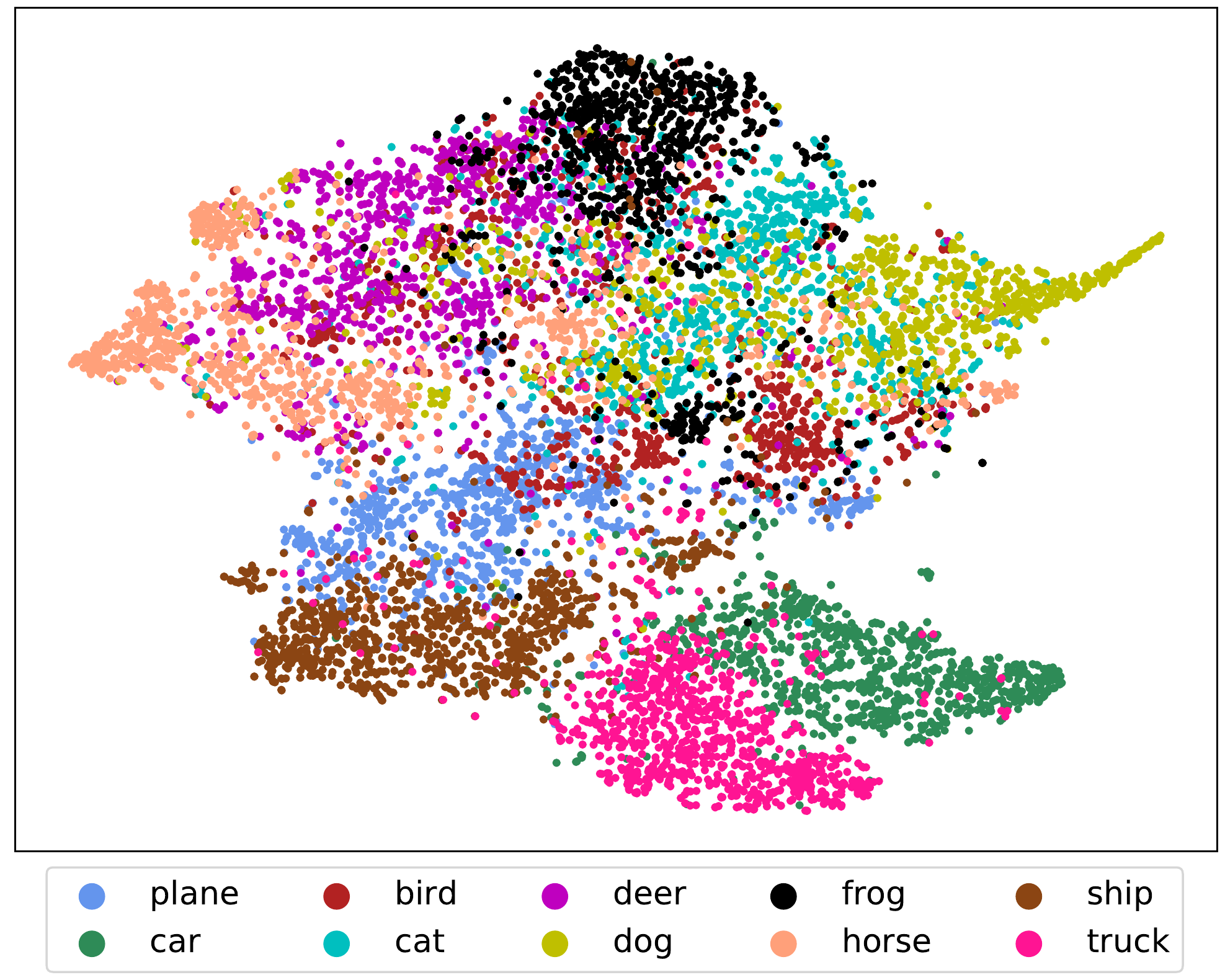}}
    \subfigure[FedU on Non-IID Data]{\label{fig:tsne-fedu-noniid}\includegraphics[width=40mm]{charts/tsne_whole_0.2.pdf}}
    \subfigure[FedU on IID Data]{\label{fig:tsne-fedu-iid}\includegraphics[width=40mm]{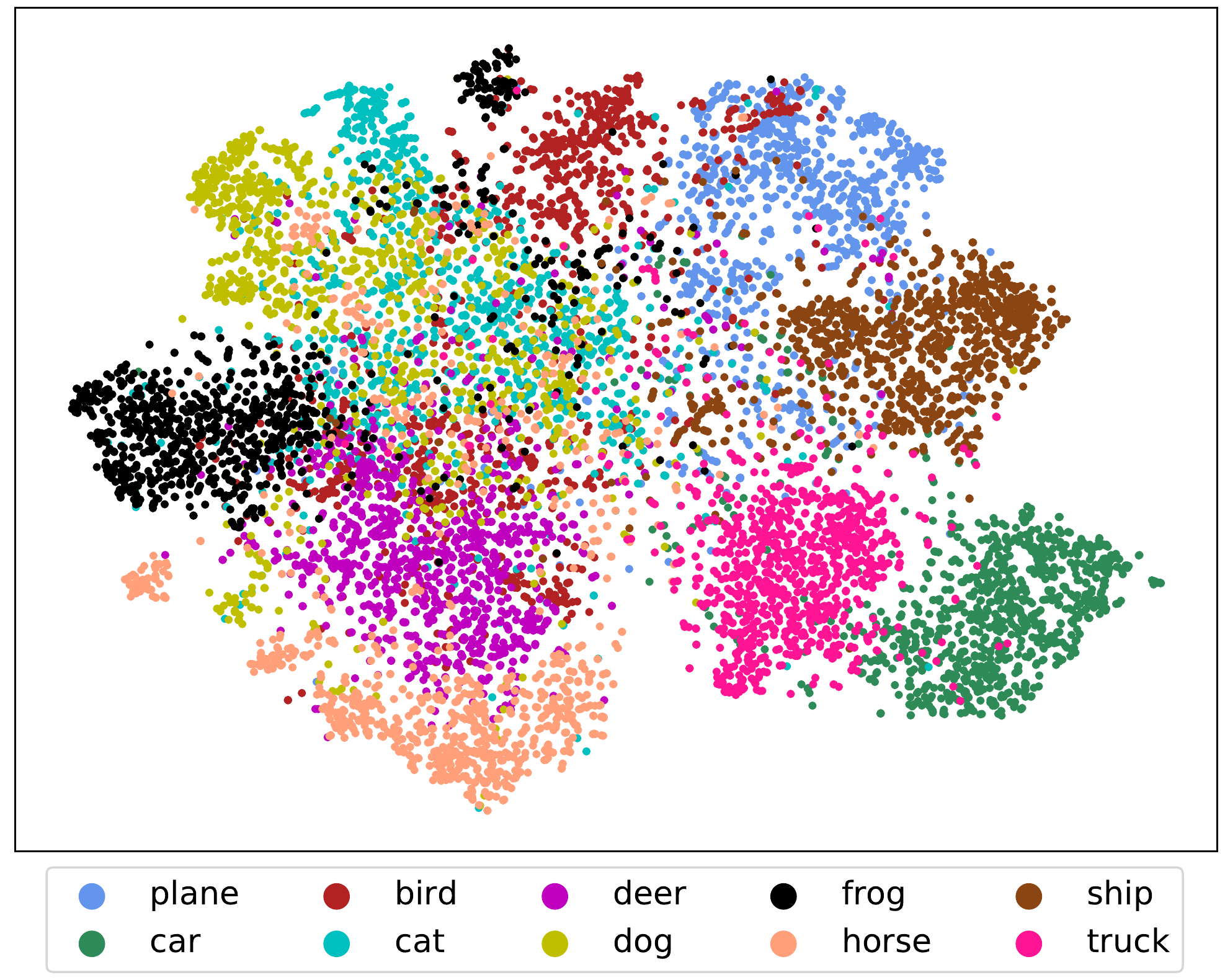}}
    \subfigure[BYOL \cite{grill2020byol} (Centralized)]{\label{fig:tsne-byol}\includegraphics[width=40mm]{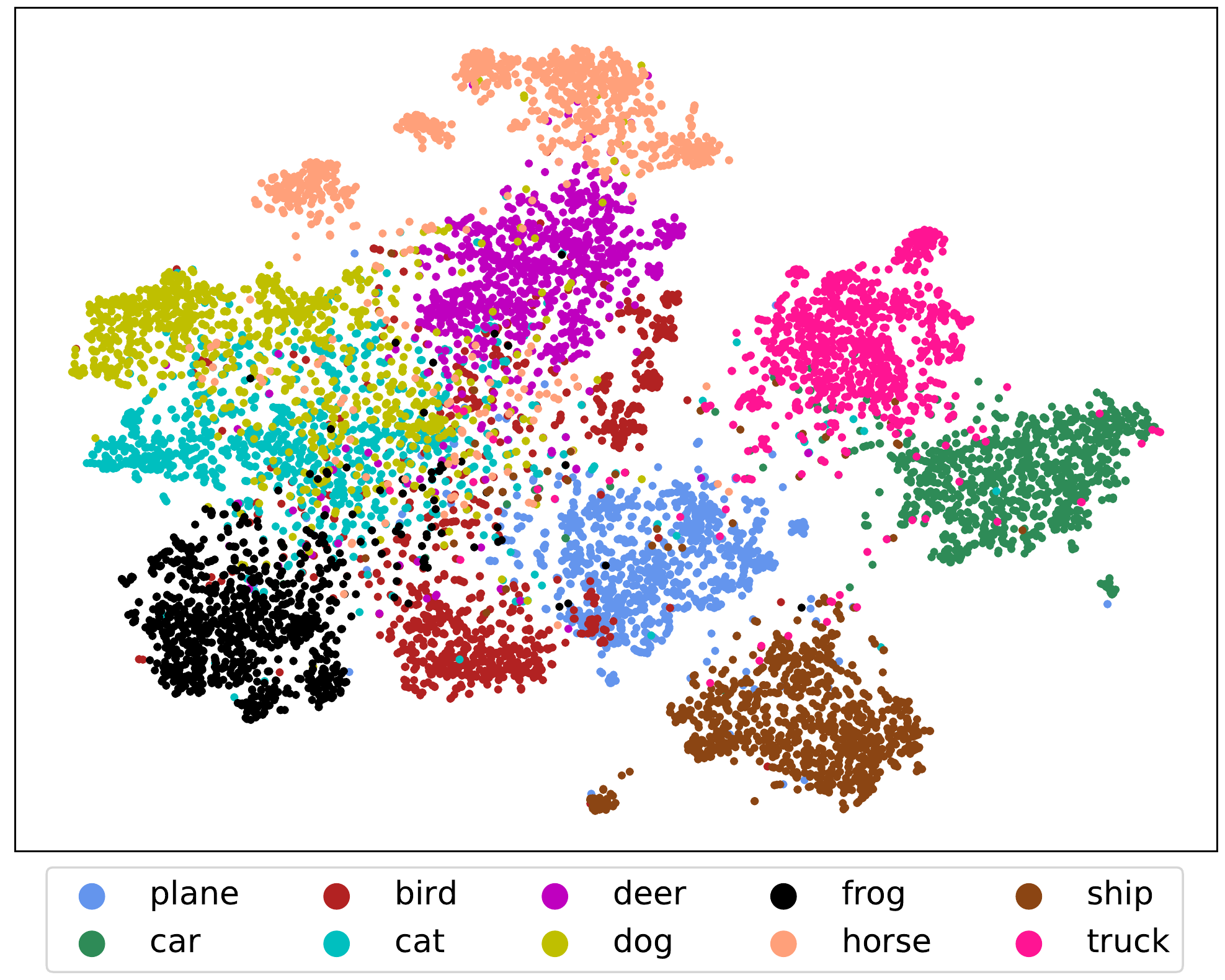}}
  \end{center}
    \caption{T-SNE visualization of representations learned from different methods: (a) Aggregate the online encoder and update the target encoder; (b) Aggregate the target encoder and update the online encoder; (c) Aggregate and update the online encoder; (d) Our proposed FedU trained on non-IID CIFAR-10 data; (e) FedU trained on IID data; (f) Centralized unsupervised learning method (BYOL \cite{grill2020byol}). (a), (b), and (c) always use the global predictor, while (d) uses DAPU to dynamically update the predictor. FedU with DAPU (d) presents better clustering results than (a), (b), and (c). FedU's representation learned from IID data (e) is also comparable with centralized training (f).}
    \label{fig:tsne-pa}
  \end{figure*}

\begin{figure}[t]
\begin{center}
  \includegraphics[width=0.6\linewidth]{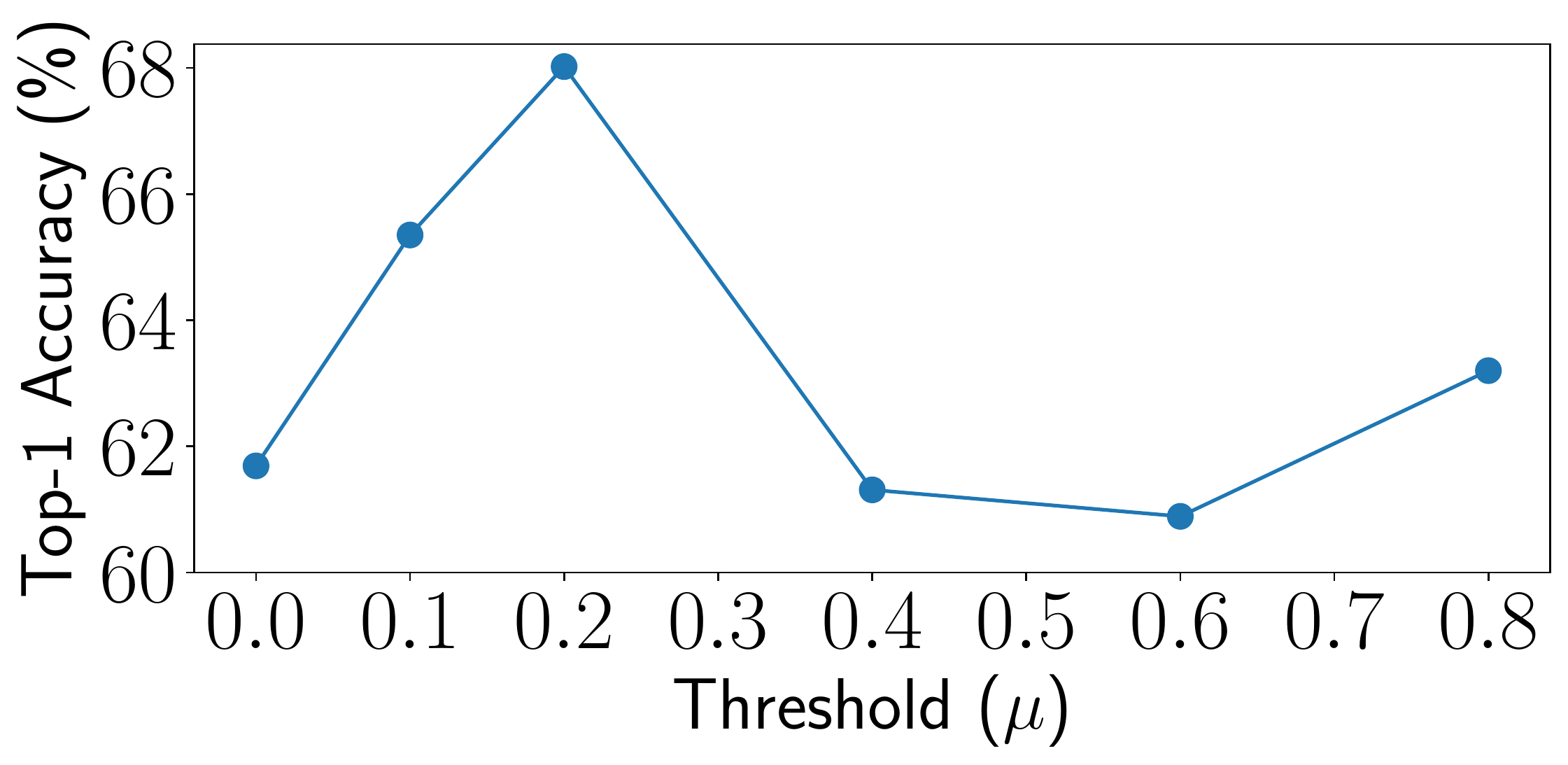}
\end{center}
    \caption{Impact of threshold $\mu$ on the non-IID setting of the CIFAR-100 dataset. DAPU with $\mu = 0.2$ achieves the best performance, complementing our results in the main manuscript.}
\label{fig:threshold-cifar100}
\end{figure}

\begin{table}[t]
\begin{center}
 \begin{tabular}{ccc} 
 \hline
 Batch Size & Adjusted LR & Constant LR \\
 \hline
 32 & 81.03 & 74.92 \\ 
 \hline
 64 & 81.69 & 79.59 \\
 \hline
 128 & 81.70 & 81.70 \\
 \hline
 256 & 81.80 & 83.22 \\
 \hline
\end{tabular}
\end{center}
 \caption{Impact of batch size on performance. A larger batch size leads to better performance when the learning rate is constant. The performances of various batch sizes are similar when the learning rate is adjusted accordingly.}
\label{tab:batch-size}
\end{table}

\textbf{Impact of Batch Size} We study the impact of batch size in Table \ref{tab:batch-size}. Constant learning rate (LR) means that we use the same learning rate $\eta = 0.032$ for experiments of different batch sizes. As for adjusted LR, we use learning rate $\eta = \frac{B \times 0.032}{128}$ for different batch sizes $B$. By adjusting the learning rate accordingly, the results are similar among batch sizes $B = \{32, 64, 128, 256\}$. However, when the learning rate is constant, a larger batch size leads to better performance. We use $B = 128$ in our main manuscript. It indicates that the experimental results in the main manuscript can be further improved with larger batch size. We run these experiments with $E = 5$ and $R = 100$ on non-IID setting of CIFAR-10.

\begin{figure}[t]
\begin{center}
  \includegraphics[width=0.67\linewidth]{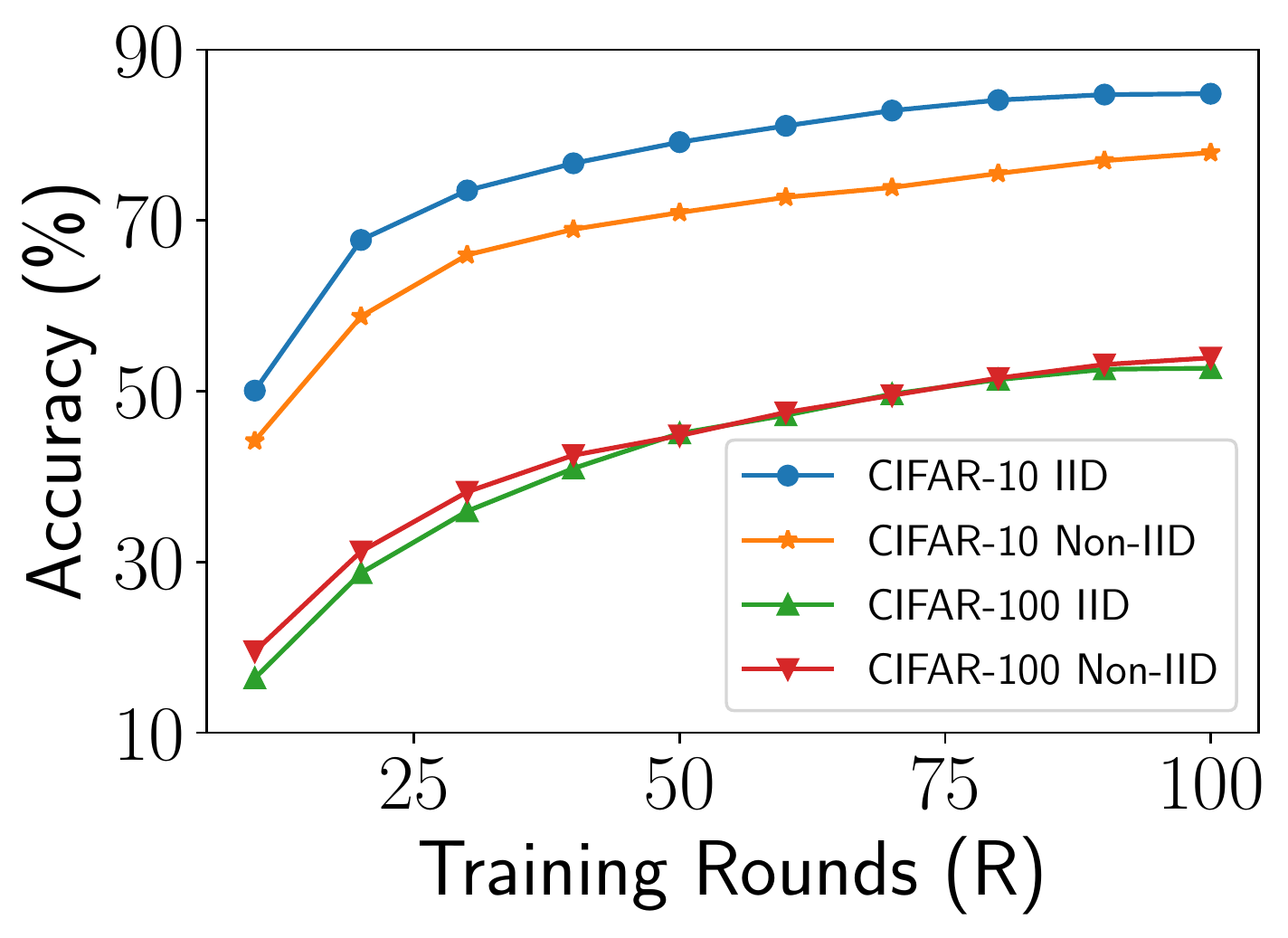}
\end{center}
    \caption{The kNN testing accuracy improves as training continues, demonstrating the convergence of FedU.}
\label{fig:convergence}
\end{figure}

\textbf{Convergence of FedU} Figure \ref{fig:convergence} shows that FedU has nice convergence property --- the accuracy steadily improves as training proceeds. We monitor the training progress by performing classification using k-nearest neighbors (kNN) \cite{wu2018npid-memorybank, chen2020simsiam}. We set the number of neighbors to 200 and the temperature to 0.1. These experiments are run with $E = 5$ and $R = 100$. 

\begin{figure}[t]
\begin{center}
  \includegraphics[width=0.67\linewidth]{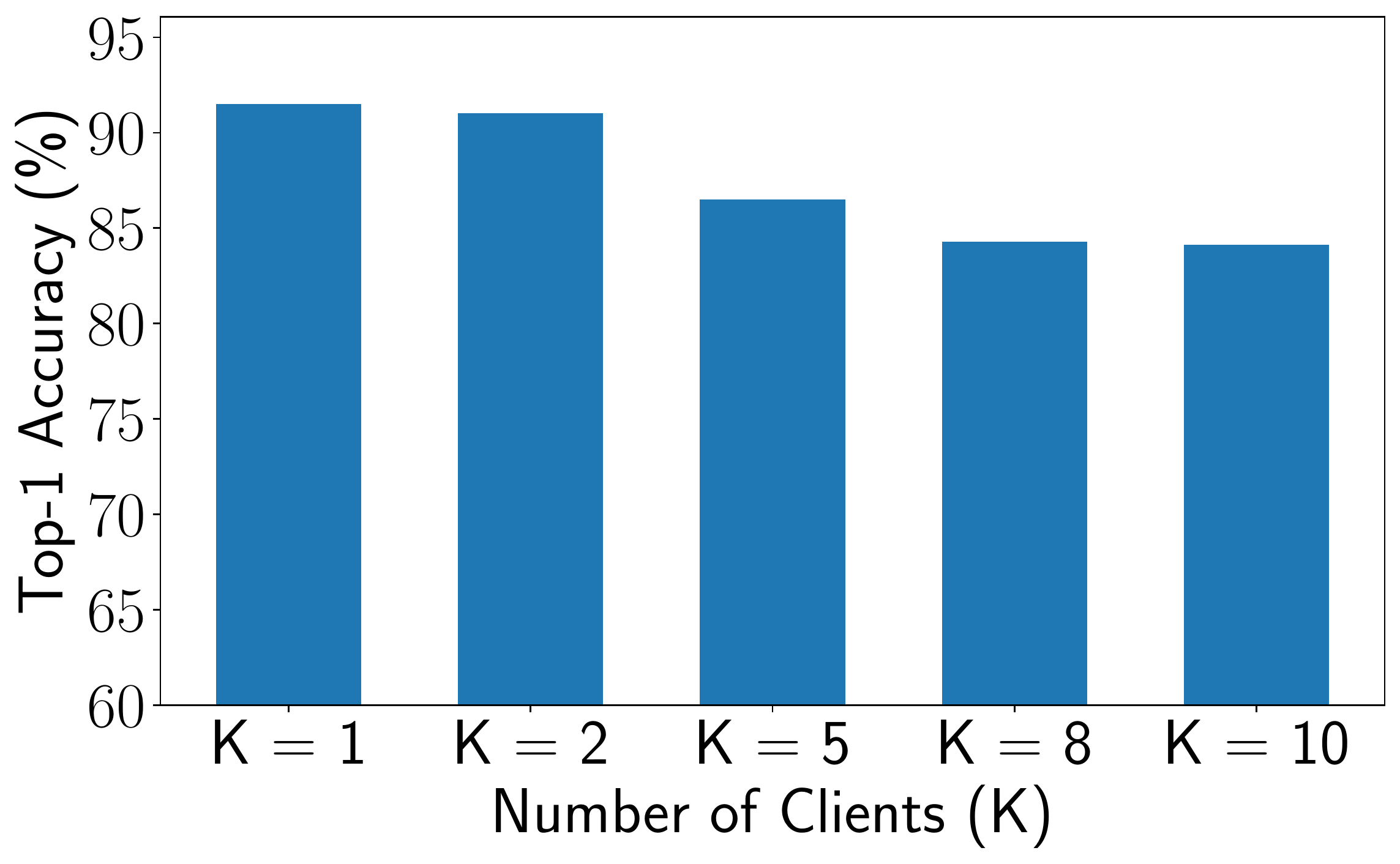}
\end{center}
    \caption{Impact of the number of clients on FedU. Although the performance decreases as $K$ increases, the performance of $K = 10$ is still better than single client training.}
\label{fig:num-clients}
\end{figure}

\textbf{Scalability of FedU} We compare the performance of different numbers of clients $K$ in Figure \ref{fig:num-clients}. We use IID setting to conduct the experiments to keep the same data distribution as we change $K$ among $\{1, 2, 5, 8, 10\}$. We split the CIFAR-10 dataset to $K$ clients with equal data volume, clients of larger $K$ contain less data. Although the performance decreases with the increase of $K$, it is still better than single client training when $K = 10$. Besides, the performance almost maintains from $K = 5$ to $K = 10$. We run these experiments with $E = 5$ and $R = 100$.

\textbf{Representation Comparison} We compare the t-SNE visualization of representations learned from different methods in Figure \ref{fig:tsne-pa}. Figure \ref{fig:online-target-global}, \ref{fig:target-online-global}, and \ref{fig:online-online-global} always uses the global predictor to update clients' predictors, complementing the t-SNE visualizations on the main manuscript where the local predictor are always used. FedU with DAPU \ref{fig:tsne-fedu-noniid} has better clustering results than the first three, indicating the effectiveness of DAPU. Besides, FedU's representation learned from IID data (Figure \ref{fig:tsne-fedu-iid}) is also comparable with centralized training (Figure \ref{fig:tsne-byol}).




{\small
\bibliographystyle{ieee_fullname}
\bibliography{egbib}
}